\documentclass[preprint]{aastex631}

\usepackage{amsmath}

\shorttitle{MCMC for Bayesian Parametric Galaxy Modeling}
\shortauthors{Buchanan et al.}

\graphicspath{{./}{figures/}}

\begin{document}

\title{Markov Chain Monte Carlo for Bayesian Parametric Galaxy Modeling in LSST}

\correspondingauthor{James J. Buchanan}
\email{buchanan11@llnl.gov}

\author[0000-0001-8207-5556]{James J. Buchanan}
\affiliation{Physics Division, Lawrence Livermore National Laboratory, Livermore, CA 94550, USA}

\author[0000-0002-8505-7094]{Michael D. Schneider}
\affiliation{Physics Division, Lawrence Livermore National Laboratory, Livermore, CA 94550, USA}

\author[0000-0002-2911-8657]{Kerianne Pruett}
\affiliation{Physics Division, Lawrence Livermore National Laboratory, Livermore, CA 94550, USA}

\author{Robert E. Armstrong}
\affiliation{Physics Division, Lawrence Livermore National Laboratory, Livermore, CA 94550, USA}

\begin{abstract}
We apply Markov Chain Monte Carlo (MCMC) to the problem of parametric galaxy modeling, estimating posterior distributions of galaxy properties such as ellipticity and brightness for more than 100,000 coadded images of galaxies taken from DC2, a simulated telescope survey resembling the ongoing Rubin Observatory Legacy Survey of Space and Time (LSST). This analysis focuses only on truly unblended galaxies detected as single objects. We use a physically informed prior, apply selection corrections to the likelihood, and systematically study the bias and calibration of our posteriors. The resulting posterior samples support rigorous probabilistic inference of galaxy model parameters and their uncertainties, even for low signal-to-noise galaxies that are often excluded from cosmological analyses. We implement the probabilistic modeling and MCMC inference using the \texttt{JIF} (Joint Image Framework) package, which we make freely available.
\end{abstract}


\section{Introduction} \label{sec:intro}
\subsection{LSST and the Need for Bayesian Galaxy Modeling}\label{subsec:intro_lsst}
The Vera C. Rubin Observatory officially started the 10 year Legacy Survey of Space and Time (LSST; \citealt{2019ApJ.873.2}) on June 29, 2026 (\citealt{rtn-011} v9.0). The LSST will observe an unprecedented number of distant galaxies throughout a wide region of the celestial sphere. A key motivation for modeling the properties of many distant galaxies is to analyze their correlations and thereby infer quantities such as weak lensing shear. Increasing the number of well-measured galaxy shapes is generally expected to improve the statistical strength of weak lensing inferences, as long as systematic errors can be controlled \citep{2018ARAA.56}.

Standard maximum likelihood estimators of galaxy shapes suffer from numerous biases, from such sources as noise \citep{2012MNRAS.425.3}, point-spread function (PSF) distortions \citep{2017AandA.604.A109}, and centroid estimation errors \citep{2017MNRAS.471.1}. In addition to its own irreducible contributions to uncertainty, noise bias interacts with and amplifies the effects of model bias, the inability of a given galaxy model to exactly fit the truth \citep{2014MNRAS.441.3}. These sources of bias must be calibrated away using estimator-specific methods (e.g., \citealt{2017MNRAS.471.1}), which may still leave behind systematic uncertainties that are not always well understood.

In contrast, a Bayesian forward modeling approach need not be similarly subject to the biases noted above: the noise level and PSF, and other effects on image rendering, can in principle be forward-modeled and thus naturally accounted for without a separate calibration step. Galaxy shape uncertainties can be described in a Bayesian sense by selecting a parametric family of galaxy light profiles, asserting a prior probability distribution on the profile parameters, and then finding the posterior probability distribution over these parameters for any specific observed data instance.

Methods such as \textsc{Metadetection}~\citep{2025MNRAS.543.4156Y, 2023OJAp.6} and FPFS~\citep{Park2025.537.1, Li2018.481.4} / \texttt{AnaCal}~\citep{Li2024.527.4} apply a perturbation-based analysis of the response of the galaxy shapes to the distorting field, which mitigates noise and selection biases while avoiding the systematic uncertainties introduced by ad hoc calibrations~\citep{Li2022.511.4}. These have been shown to produce estimates of cosmic shear with systematic uncertainties satisfying stage-IV survey requirements~\citep{2023OJAp.6, Li2025.jun}. Nevertheless, a point estimate of any value, whether that be galaxy ellipticity or cosmic shear, even when accompanied by a confidence interval, fails to capture all the information available in one's limited dataset. A Bayesian approach can, at least in principle, describe in precise distributional form the full state of our knowledge (for some assumed prior).

Starting from sampled Bayesian estimates of individual galaxy properties, hierarchical methods guided by importance sampling may be used to infer Bayesian estimates of derived values such as cosmic shear and cosmological parameters. \citet{2015ApJ.807.1} outlined a particular scheme for doing so. The present study describes an early stage of that hierarchy. Spatially correlated shear-dependent selection effects, for example, need not be accounted for at this step, but rather appear in subsequent stages of hierarchical inference. Section~\ref{sec:footprints} discusses earlier steps in the process involving object detection and characterization, and Section~\ref{sec:conclusions} sketches their connection to galaxy modeling as part of a holistic inference pipeline.

\subsection{Efforts toward Bayesian Galaxy Modeling}
The premise of computing the likelihood via comparing simulated galaxy image models to real images was laid out in \citet{1999AandA.352} as a way of potentially improving on the Kaiser--Squires--Broadhurst \citep{1995ApJ.449} method for estimating weak lensing shear. This forward image modeling approach was incorporated into Bayesian posterior inference in \texttt{Im2shape} \citep{2001Bridle, 2013ascl.soft07006B}, which was tested alongside other weak lensing estimation methods as part of the Shear Testing Programme (STEP; \citealt{2006MNRAS.368.3}).

STEP used SExtractor \citep{1996AAPS.117} to supply the initial detections for all methods. Here, we instead use detections provided by the footprint construction and peak finding algorithms of the LSST Science Pipelines, described in Section~\ref{sec:imagesim}. Our rationale for using Pipelines detections is twofold. First, it is convenient, since a large set of those have already been made and publicly shared, with corresponding truth information, as part of DC2 and Rubin Observatory's Data Preview 0.2.\footnote{\href{https://dp0-2.lsst.io/}{\nolinkurl{dp0-2.lsst.io}}} More significantly, something close to this exact detection procedure is likely to be implemented in early LSST data releases, making our analysis of more direct relevance to LSST science than one based on SExtractor.

Whereas \texttt{JIF}, \texttt{Im2shape}, and other methods discussed below use Markov Chain Monte Carlo (MCMC) for posterior estimation, the \texttt{LensFit} \citep{2007MNRAS.382.1} method instead marginalizes over several model parameters with a combination of analytic and numeric methods, and ultimately evaluates the entire likelihood surface in ellipticity. While this approach is appealing for ellipticity measurements due to its speed, it explicitly makes no attempt to measure the other ``uninteresting'' parameters, and it relies on significant assumptions on the specific forms of the prior and likelihood in order to make the marginalization tractable. With MCMC the closest corresponding assumptions are simply that the prior and likelihood can be computed reliably and quickly, which in principle allows more realistic forms for both.

In all of the preceding methods (including \texttt{JIF}), the basic surface brightness profile shape for any particular fit can be freely chosen: the initial \texttt{LensFit} studies focused on an exponential profile \citep{2007MNRAS.382.1, 2008MNRAS.390.1}. For a single-component light profile of this sort, the parameters needed to specify a given galaxy's top-of-the-atmosphere appearance in a single photometric band are its total brightness, some measure of its overall size such as the half-light radius (HLR), two real numbers describing its lensed ellipticity, and two numbers specifying its position on the sky. We model precisely this set of six parameters for each galaxy morphology, just as in the \texttt{LensFit} studies, though we consider de Vaucouleurs (bulge) alongside exponential (disk) light profiles. In addition, we model three parameters for the PSF, describing its overall size and ellipticity, for a total of nine continuous fit parameters for each of the two light profile types.

\subsection{MCMC for Bayesian Galaxy Modeling}
A number of Bayesian parametric galaxy modeling codes have been developed in recent years including \texttt{ProFit} \citep{2017MNRAS.466.2}, \texttt{PyAutoGalaxy} \citep{2023JOSS.8.81}, \texttt{pysersic} \citep{2023JOSS.pysersic}, \texttt{BANG} \citep{2023MNRAS.525.1}, and \texttt{AstroPhot} \citep{2023MNRAS.AstroPhot}. All of these implement one or more variants of MCMC. Given the many tools in this ecosystem and the evident demand behind them, it is worth systematically examining how this approach performs across a physically plausible range of inputs with realistic complications. In this paper, we study MCMC for Bayesian parametric galaxy modeling in the specific context of distant galaxies detected by LSST.

If the model family contains light profiles sufficiently close to the true light distribution of any given galaxy, if the assumed prior is close to the overall distribution of true model parameters in the dataset, if the production of observed images of any given galaxy can be modeled sufficiently well, and if the corresponding posterior can be computed exactly, then the posterior mean will be an unbiased estimator of the model parameters \citep{2007MNRAS.382.1}. In practice, the model family, the assumed prior, the assumed likelihood, and the computed posterior will all be approximations at best.

In order to study how well our approach can recover true underlying galaxy properties, as well as to examine it in the specific context of LSST, we use images and associated data products from the DC2 simulated sky survey \citep{2021APJS.253.1}. By construction, the galaxies in these images have realistically correlated distributions of apparent sizes, morphologies, luminosities, and positions in the sky, and are rendered as they would be perceived by LSST.

As in most previous studies of Bayesian parametric galaxy modeling (e.g., \citealt{2007MNRAS.382.1, 2008MNRAS.390.1}), we confine our attention to individual galaxies that do not significantly overlap others and are well-separated from their nearest neighbors. Due to the extreme depth of LSST, such totally ``unblended'' galaxies will be the exception rather than the rule \citep{2021NatRevPhys.3}, further motivating the study of galaxy model fitting methods in this specific survey context. This study of isolated galaxies provides a baseline for comparing the performance of our approach with others, along with a foundation for future blending studies. The extension of the current method to more general images including blending is briefly discussed in the concluding section.

In this study, we fit more than 100,000 isolated DC2 galaxies using an MCMC method. In each iterative step of an MCMC chain for a given galaxy, we sample some assumed galaxy parameters, render a simulated image of a galaxy light profile with those parameters, and compare this to the observed telescope image to compute the likelihood, which is the probability density of seeing the observed image given the assumed parameters. The likelihood together with the prior probability density on the assumed parameters informs the next step of the chain. Our code for handling this is contained in the JIF (Joint Image Framework) package~\citep{jif}.\footnote{\href{https://github.com/mdschneider/JIF}{\nolinkurl{github.com/mdschneider/JIF}}} For this initial study, we confine our attention to single-band simulated Rubin images.

\subsection{Novel Developments}
Of the above-cited efforts, the most similar to our own are \citet{2017MNRAS.466.2} and \citet{2023MNRAS.525.1}, which present thorough studies of real or realistic simulated datasets consisting of 10,000 galaxies each. The \texttt{ProFit} study is particularly noteworthy here for examining simulated LSST data. Here, we examine ten times as many galaxies, with a special emphasis on ``getting things right'' for the dimmest galaxies, which are expected to comprise the vast majority of LSST detections. Real data can include such complications as masked pixels, invalid pixel values, or pathological variance plane estimates, all of which need to be accounted for in the likelihood model. \texttt{JIF} accommodates all of these and runs to completion without issues on every example in our dataset, so long as we assume a physically realistic prior on galaxy model parameters.

The present study takes the rare step, in the recent literature on Bayesian galaxy modeling in deep optical sky surveys, of explicitly formulating a specific, highly informative prior. A noteworthy exception is the study of \citet{2023MNRAS.525.1} applying \texttt{BANG} to galaxies in the Sloan Digital Sky Survey (SDSS)-MaNGA survey, in which they considered three different priors of varying strength. To study the potential benefits of using an informed prior, we adopt a prior that matches as closely as possible the true distribution of galaxy properties in our dataset.

We also take the novel step, in Bayesian parametric galaxy modeling studies, of developing selection corrections to the likelihood. These are required due to the fact that all data examples we examine must pass a significant set of selection criteria to be considered, which means that the set of data we see is systematically skewed from what a more naive, unconditioned likelihood model would predict. As we show, both the prior and the selection corrections play dominant roles in determining the posteriors for low signal-to-noise ratio (S/N) galaxies, which are expected to comprise a large fraction of the total number of galaxies detected in LSST.

Unlike the SDSS-MaNGA dataset used in the \texttt{BANG} study, we fit here many dim galaxies with true average brightness levels well below the threshold for detection, which enter our dataset through a chance accumulation of noise in the DC2 images. This is a major reason we adopt such a simple description of galaxy morphology, as opposed to, e.g., the 18-parameter model used in \citet{2023MNRAS.525.1}. Our parameterization suffices to capture much of the morphological range in the DC2 simulations, and the minimal number of parameters simplifies the inference, though \texttt{JIF} is designed to readily incorporate arbitrary parametric models. For example, functionality already exists in \texttt{JIF} for modeling a generic Sérsic index.

Section~\ref{sec:imagesim} reviews how the DC2 images were simulated and describes how galaxies were detected and selected for this study. Section~\ref{sec:parameters} defines the model parameters we fit. Section~\ref{sec:prob_model} introduces the probabilistic data model we assume for fitting, while Section~\ref{sec:selection_corrections} discusses selection corrections to the likelihood. Section~\ref{sec:mcmc} details our specific MCMC fitting procedure. The end product of MCMC is a sequence of model parameter values, which ideally comprise a statistically fair sample from the posterior. Section~\ref{sec:results} examines the fidelity of this probabilistic interpretation. We conclude in Section~\ref{sec:conclusions} where we outline what remains to be done to incorporate this method into a complete Bayesian inference pipeline for LSST data analysis.

\section{Simulated Survey Data}\label{sec:imagesim}
\subsection{The DC2 Survey}
The data we use comes from a set of simulated images of a simulated galaxy distribution produced by the LSST Dark Energy Science Collaboration (LSST DESC; \citealt{2012_DESC}).\footnote{The description of DC2 in this section borrows substantially from \citet{2022ApJ.924.2}.} Galaxies in the simulated distribution have properties described by the CosmoDC2 v1.1.4 catalog \citep{2019APJS.245.2} and the DESC DC2 Truth Match dr6 v1 catalog \citep{2021_DESCDC2_datarelease}. The distribution of galaxies in the CosmoDC2 catalog is based on the Outer Rim run \citep{2019APJS.245.1}, a trillion-particle, $(4.225\,\text{Gpc})^3$ simulation using a cosmological model based on Wilkinson Microwave Anisotropy Probe data \citep{2011APJS.192.2}. The catalog contains a realistic number density of galaxies up to a depth of $z = 3$ and an \textit{r}-band magnitude of 28, along with many additional galaxies at higher magnitudes. Numerous CosmoDC2 catalog properties are in close agreement \citep{2019APJS.245.2} with observed data from a variety of sources including the HSC \citep{2018PASJ.70.SP1.b}, COSMOS \citep{2007APJS.172.1}, DEEP2 \citep{2013APJS.208.1}, and SDSS \citep{2017APJS.233.2} surveys.

CosmoDC2 is the observing target of the LSST DESC DC2 Simulated Sky Survey \citep{2021APJS.253.1}, a simulation of LSST observations covering a $300\,\text{deg}^2$ region of the southern sky, with a total number of exposures corresponding to half of the total survey duration (i.e., 5 years). The DESC DC2 Truth Match catalog contains the true values of certain measurable properties of galaxies imaged in the survey, including their total fluxes integrated over each filter bandpass, accounting for the effects of internal reddening, redshift, reddening from Milky Way dust, and atmospheric and instrumental throughputs. For the purpose of defining ``true'' parameter values, the flux as well as the gravitationally lensed R.A.\ and decl.\ for each galaxy were taken from this catalog. Other quantities used to compute true parameters --- the weak lensing convergence and shear components, overall weak lensing magnification, unlensed ellipticity components, and unlensed half-light radius for each galaxy --- were taken from the CosmoDC2 catalog.

The major data product from the DC2 Survey is a set of simulated telescope images that closely model how galaxies in the CosmoDC2 distribution would be seen by LSST. The light from each galaxy is modeled as the sum of a Sérsic bulge ($n = 4$) and disk ($n = 1$), plus some random Gaussian-distributed local flux excesses representing ``knots'' of star formation. This light is modeled as it travels through intergalactic space, Milky Way dust, Earth's atmosphere, the telescope optics, and finally through collection and digitization by the telescope CCDs. In addition to galaxies, the DC2 images include sky background, Milky Way stars, simulated Type Ia supernovae, and certain technical artifacts of the simulated camera instrumentation and pointing.

\subsection{Pipelines Processing}
The raw DC2 images have been processed by the LSST Data Management Science Pipelines software stack (LSST Science Pipelines)\footnote{\href{https://pipelines.lsst.io}{\nolinkurl{pipelines.lsst.io}}}, a unified software suite for astronomical image data processing that takes care of many tasks including instrument signature removal, calibration, background subtraction, variance and PSF estimation, coaddition of multiple exposures into signal-boosted composite images (\textit{coadds}) of specified sky regions, and object detection and measurement. In this study, we exclusively use \textit{i}-band coadd images.

For the purpose of coaddition, the DC2 skymap divides the sky into partially overlapping \textit{tracts}, each of which is further subdivided into a $7\times7$ grid of roughly square \textit{patches}. Each patch corresponds to one coadd image. The specific coadds we use correspond to each of the 49 patches that form tract 3830 of the DC2 skymap. The interior region of each patch consists of a $4000\times4000$ square grid of pixels, each of which is 0.2\arcsec\ to a side, and interior regions of patches in a tract tile the sky without gaps or overlaps. This gives one tract a total sky coverage of 2.4 square degrees, with the interior region of one patch covering 180 square arc minutes.\footnote{Each patch also has a border region extending 100 additional pixels outside its interior region, except on tract boundaries. These borders overlap with a small part of the interior regions of neighboring patches.} For this study, we associate a given galaxy to a specific patch if the true gravitationally lensed location of its center sits in the interior region of that patch. This ensures that galaxies are not duplicated in our set of footprints (described in Section~\ref{sec:footprints}) and that the entire images of galaxies are contained entirely in their corresponding patches.

\subsection{Footprints}\label{sec:footprints}
For us to model a galaxy's parameters, its existence in a given patch of space must first be asserted, and the specific pixels described by our probabilistic model must be specified. For these purposes, we use \textit{footprints}, contiguous regions of relatively bright image pixels, created by the LSST Science Pipelines as part of their procedure for object detection. First, an initial set of preliminary footprints is constructed by identifying pixels at which the point-source S/N (as defined in \citet{2018PASJ.70.SP1.a}, with an additional correction for correlated noise) exceeds 5. Each footprint is then expanded to include all pixels up to three or four pixels away from every preliminary pixel, using a diamond-shaped expansion kernel roughly 2.4 times the radius of the coadded patch's estimated PSF. Some expanded footprints may overlap, so any overlapping cases are finally merged into single footprints. A footprint's \textit{bounding box} is the minimal rectangular image containing the footprint pixels.

We say a footprint contains a galaxy if that galaxy's lensed R.A.--decl.\ position listed in the Truth Match catalog sits inside one of the footprint's pixels (as associated to R.A.--decl.\ by the DC2 skymap). We define a footprint to be truly ``unblended'' if the brightest galaxy contained in the footprint has a lensed, Milky Way-reddened, \textit{i}-band total nJy flux (what we generally mean by ``flux'' from here on) at least 50 times greater than the fluxes of all other galaxies contained in the footprint summed together, again using the values listed in the Truth Match catalog.

The Pipelines estimate the number of galaxies in a footprint by asserting that they are in one-to-one correspondence with local brightness peaks. Peaks are found by convolving the image with a Gaussian approximation of the average PSF in a patch, and then identifying pixels in the convolved image with intensities at least as large as all eight of their immediate neighbors. Temporary local background oversubtraction is applied at this stage to suppress the number of spurious peaks that can arise in large footprints. Peak finding is performed independently in each of the six filter bands, and the separate peak collections are finally merged together by associating peaks across different bands if they are sufficiently close, described more fully in \citet{2018PASJ.70.SP1.a}.

If a footprint is too close to the edge of a patch, then the PSF convolution used for detection becomes unreliable. We ignore footprints that are closer to the edge of the patch than the width of the PSF convolution kernel, which here amounts to 13 pixels. To avoid double-counting, we only model footprints whose brightest galaxies sit inside the patch's interior region (recall these interiors tile the tract without overlapping). As we show in sec.~\ref{sec:prior}, the true location of an unblended galaxy almost always lies within 3 pixels of the center of the footprint bounding box, so we would get essentially the same selection criterion by requiring the bounding box center to lie at least 3 pixels inside the patch interior.

A small fraction of pixels in any given coadd will be affected by nonastrophysical, non-sky-background sources of flux variation, such as cosmic rays or electronic issues. We ignore any footprints that have any pixels flagged by the Pipelines with any of the following known issues: ``BAD'', ``SAT'' (saturated), ``INTRP'' (interpolated), ``CR'' (cosmic ray), ``SUSPECT'', ``CLIPPED'', ``CROSSTALK'', or ``UNMASKEDNAN''. Finally, a fraction of unblended footprints contain stars or supernovae. For simplicity's sake, we ignore any such footprints. For a more realistic analysis, one could run a machine learning-based (ML) star vs.\ galaxy classifier on the footprints as in \citet{2022ApJ.163.4} and only select those having galaxy scores above some threshold with a desired purity.

We will say that truly unblended single-peak footprints are ``isolated.'' Because each isolated footprint has exactly one true galaxy with overwhelming relative brightness, it is straightforward to compare the results of a single-component fit to corresponding true parameters. It is often the case that any given footprint with one peak is truly unblended and vice-versa, but just as often there are other galaxies in the footprint with at least $1/50$th the total flux of the brightest galaxy. As with star vs.\ galaxy classification, blend identification can potentially be done more reliably with an ML classifier, on which one can set a tunable threshold~\citep{Liang2026.9, 2022ApJ.924.2}.

A typical coadd (patch 40 in tract 3830) contains 116,406 galaxies, 650 stars, and 74 supernovae, producing 12,737 total detection footprints (after merging across all 6 filter bands), 7929 of which have exactly one peak. Of those 7929, 151 are not well-contained in the patch region, and 175 have a flagged pathological pixel in the \textit{i}-band. With such examples excluded, 7606 footprints remain, of which 82 contain a star, 5 contain a supernova, and 53 do not contain any galaxies. Of the remaining 7492 unpolluted footprints, only 3388 meet our definition of ``truly unblended,'' and 3151 of these have their galaxy located within the patch interior. Most of the 49 patches we consider similarly have around 3000 footprints that pass all of our selection criteria.

The precise definition of ``blended'' significantly affects the sample. For our very strict $1/50$ maximum on the fraction of blended flux, the selection efficiency on good single-peak, unpolluted footprints is $45\%$. By increasing the maximum blended flux fraction to $1/10$, the selection efficiency rises to $55\%$, and for $1/5$ it goes up to $66\%$. Adjusting the blend fraction definition would affect the efficiency and purity of any blend classifier one might develop, and ideally, one would tune this against the reliability of MCMC inference to maximize the statistical power of downstream measurements. For this study, we are most concerned with the faithfulness we can achieve with MCMC inference, and we are not overly concerned with maximizing selection efficiency---the results in Section~\ref{sec:results} appear to be backed by a reasonable sample size.

We only fit isolated footprints that pass all of the selection criteria above, using a parametric model of a single galaxy. Several of these criteria directly make use of known truth information: whether a footprint is polluted by a star or supernova; whether the galaxy center is truly contained in the patch interior; and most significantly, whether the footprint is truly unblended. As noted above, proxy criteria for these with varying degrees of efficiency and purity can be developed based on the images alone, but as a point of departure, we concern ourselves here with what can be recovered in principle given some highly reliable knowledge of the truth. In total, 143,143 footprints in our dataset pass all the selection criteria above. We fit about four-fifths of these, as described in the next section.

For the purpose of image modeling during MCMC sampling, we must convert an assumed true flux value in nJy to an image pixel value (\textit{instrumental flux}). We perform this conversion based on a linear fit of Truth Match catalog flux vs.\ the instrumental flux in isolated footprints. See Appendix~\ref{sec:photometric} for further details of this fitting procedure. The Pipelines include an independent estimate for the photometric calibrations, which is used to good effect in synthetic source injection (SSI; \citealt{Anbajagane2025Dark}). We developed our own here based on known truth values in an effort to see how closely we could fit the flux of DC2 galaxies if we could eliminate photometric calibrations as a plausible source of error, an issue we return to in Section~\ref{sec:realistic_results}. 

\section{Model Parameters}\label{sec:parameters}
In this study, we fit a model for each isolated footprint with nine continuous parameters. Six of these describe the astrophysical light profile of the galaxy itself: its half-light radius in arcseconds, total flux in nJy, the $x$- and $y$-position of the galaxy centroid in pixel coordinates, and two values describing its sheared ellipticity. The remaining three parameters describe the PSF: its FWHM in arcseconds, plus two values for its ellipticity. We compare each of these values to associated true values. We take the true values for the six galaxy parameters from the DC2 catalogs. The lensed flux and position are provided in the DC2 Truth Match catalog, but only unlensed values are provided for half-light radius and ellipticity. Therefore, we compute their lensed values ourselves, using lensing information provided in the catalogs.

The lensed half-light radius is the unlensed value multiplied by the square root of the lensing magnification. The $x$- and $y$-components of the lensed ellipticity are the real and imaginary components, respectively, of the expression
\begin{equation}
\label{eq:e}
e = \frac{e^{i} + g}{1 + g^{*}e^{i}}
\end{equation}
where $e$ is the (complex) lensed ellipticity, $e^i$ is the object's intrinsic ellipticity, and $g$ is the reduced shear \citep{2001PhysRep.340} defined by
\begin{equation}
\label{eq:g}
\begin{gathered}
g = \gamma / (1 - \kappa), \\
\gamma = \gamma_{1} + i\gamma_{2}.
\end{gathered}
\end{equation}
These equations\footnote{As noted in \citet{2001PhysRep.340}, eq.~\ref{eq:e} for $e$ applies in general only when $|g| \leq 1$. This is true for all galaxies in our dataset.} hold for the ellipticity convention used by \texttt{JIF} for the parameters $e1$ and $e2$, and correspond to the ``reduced shear'' convention in \texttt{GalSim} \citep{2015AandC.10}. The CosmoDC2 extragalactic catalog includes true values for the intrinsic ellipticity $e^{i}$ plus the lensing $\gamma_{1}$, $\gamma_{2}$, and $\kappa$ values for each galaxy.

While each DC2 galaxy is allowed in principle to be the sum of distinct bulge and disk components, the vast majority of galaxies in the CosmoDC2 catalog are overwhelmingly dominated by one or the other component. For simplicity, we assume that the overall light profile type for every galaxy is either that of a pure bulge (de Vaucouleurs profile, Sérsic $n = 4$) or pure disk (exponential profile, Sérsic $n = 1$), with no random Gaussian knots of star formation. We assign a type label to each CosmoDC2 galaxy by first rendering its true bulge-plus-disk light profile based on catalog truth values, and then selecting whichever simplified type, bulge or disk, gives the highest likelihood to the combined bulge-plus-disk profile when everything else about the morphology (half-light radius, ellipticity, and position) matches the truth.\footnote{For this purpose we simulate extremely bright profiles having an S/N orders of magnitude higher than we expect to see in practice, to suppress the influence of background noise in the likelihood.} We declare a cosmoDC2 galaxy to be ``very bulge-like'' if its best type label is bulge, and if furthermore at least 80\% of its flux belongs to the bulge component of its true bulge+disk model; similarly for very disk-like. Out of all our isolated footprints, 81\% have a very bulge-like or very disk-like galaxy, while the remaining 19\% have a more even bulge+disk combination.

It is in fact straightforward to include the Sérsic $n$ value as an additional fit parameter, or an equivalent index in a similar parametric model such as a Spergel profile \citep{2010ApJS.191}. Preliminary experiments suggest that a posterior for such a parameter can, in principle, be modeled in conjunction with the other parameters, given a suitable prior. However, it is somewhat difficult to fully model it here in a way that our MCMC procedure can reliably handle, because the true distribution of CosmoDC2 galaxy properties is strongly bimodal, with one mode for bulges and the other for disks. Meanwhile, real galaxies do not always adhere well to either a one-component Sérsic profile or a two-component bulge-plus-disk profile. We leave investigation of this substantial topic to future work. For now, in addition to the selection criteria listed in Section~\ref{sec:footprints}, we only fit the 81\% of footprints with true galaxies that are very bulge-like or disk-like. This finally leaves us with 116,088 footprints that we fit using MCMC.

The net PSF in the DC2 simulations includes the effects of both a simulated multilayer atmosphere and the simulated telescope optics. Here, we reduce that complexity down to a three-parameter fit describing the FWHM and ellipticity of a single-component Kolmogorov PSF.\footnote{See \citet{2015AandC.10} for the specific functional forms of the exponential (disk), de Vaucouleurs (bulge), and Kolmogorov profiles used in this study.} The Pipelines processing for DC2 included PSF estimation on the coadds, supplying a corresponding PSF image at any pixel location on the coadd, which can be convolved with a galaxy light profile for, e.g., SSI. We assigned ``true'' Kolmogorov parameters to these Pipelines-estimated PSF images, using a similar procedure for how we assigned true type labels: we simulated a GalSim image of an extremely bright galaxy with known true shape parameters, convolved with the Pipelines-estimated PSF image at the known true galaxy location, and found the maximum-likelihood fit of an elliptical Kolmogorov PSF convolved with the same galaxy model. For such fits, we use the likelihood model described in sec.~\ref{sec:likelihood} and follow the same general procedure described in sec.~\ref{sec:mcmc} for maximum a posteriori (MAP) finding.

\section{Probabilistic Model}\label{sec:prob_model}
\subsection{Prior}\label{sec:prior}
The chief aim of this study is to compute the posterior probability distribution of the parameters of our model for the dominant galaxy in each isolated footprint. We begin by provisionally assuming each footprint can be fit independently. Correlated shape measurements can arise from such influences as PSF correlations~\citep{guerrini2024}, cosmic shear~\citep{li2023hypersuprimecamyear3}, and intrinsic alignments~\citep{Lamman2024IA}, but we do not consider those field-level properties at this stage. Following the program of \citet{2015ApJ.807.1}, later stages of hierarchical inference can combine the separate galaxy posterior samples from this initial stage into a common posterior on field-level properties (more on this in Section~\ref{sec:conclusions}).

We assert a prior probability distribution on the parameters of each galaxy by approximately modeling the true distribution of those parameters in our dataset. Our prior assumes that ellipticities are close to uniformly randomly oriented, as well as the polar angle of a galaxy's centroid with respect to the footprint bounding box center. We observe that for disks, this assumption holds quite well, but for bulges both angles display a pronounced sinusoidal shape to their densities. Appendix~\ref{sec:angular} details how we model these distributions.

We model the remaining four galaxy parameters---the distance of the centroid from the center of the bounding box in pixels, the magnitude of the ellipticity (the sum in quadrature of the separate ellipticity components), the natural logarithm of the total flux in nJy, and the natural logarithm of the half-light radius in arcseconds---using Bayesian Gaussian mixture models (BGMMs; \citealt{2006BA.1.1}) implemented in \texttt{scikit-learn}\footnote{\href{https://scikit-learn.org/stable/}{\nolinkurl{scikit-learn.org}}}, fit to the true distribution of these values among the galaxies in our isolated footprint set.

\begin{figure}
\includegraphics[width=\textwidth]{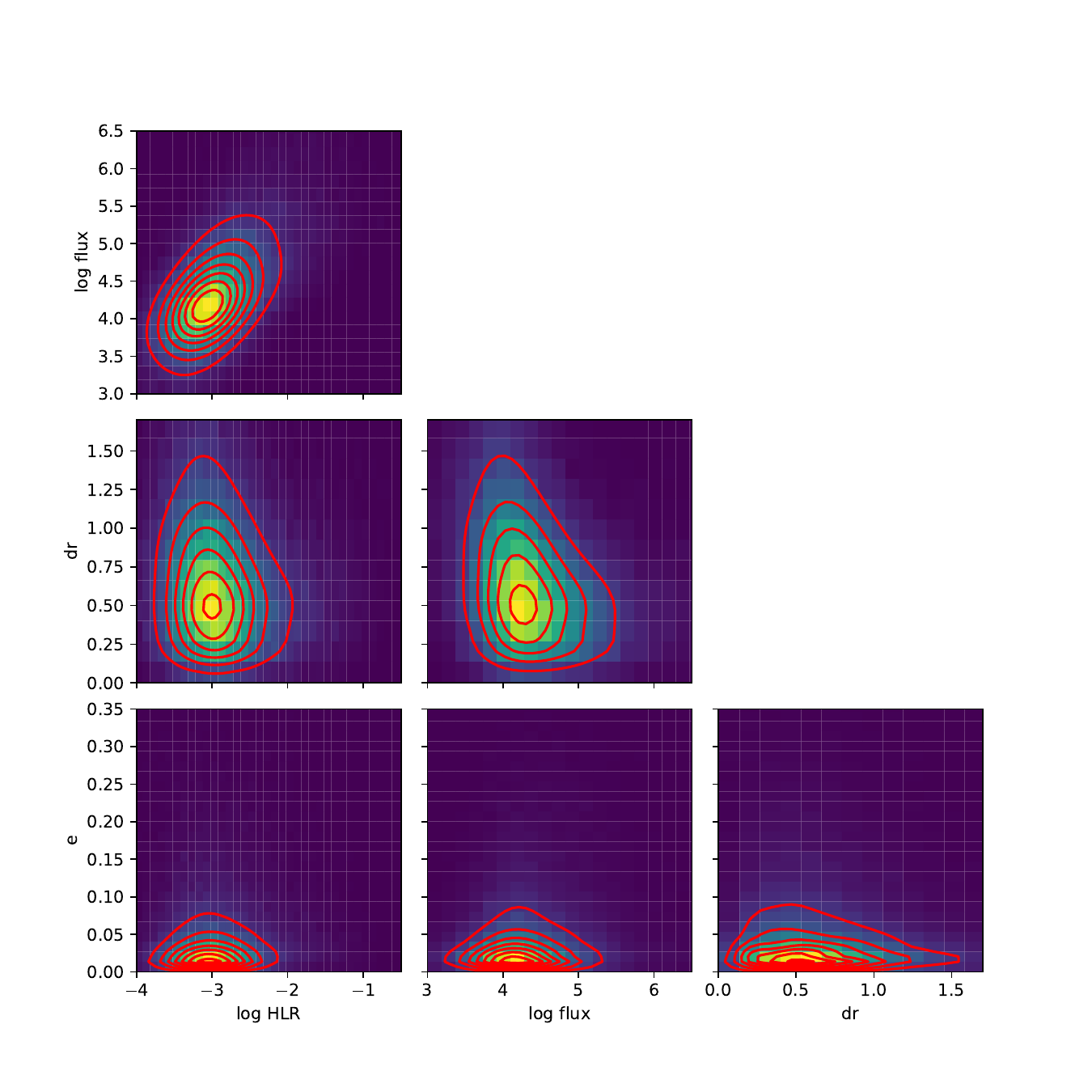}
\caption{Two-dimensional marginal densities of the BGMM model for bulges. Yellow and blue indicate histogram bins of higher and lower density, respectively. Here and elsewhere, half-light radius (HLR) is in arcseconds, flux is in nJy, and dr is in pixels.\label{fig:bulges_2d}}
\end{figure}

\begin{figure}
\includegraphics[width=\textwidth]{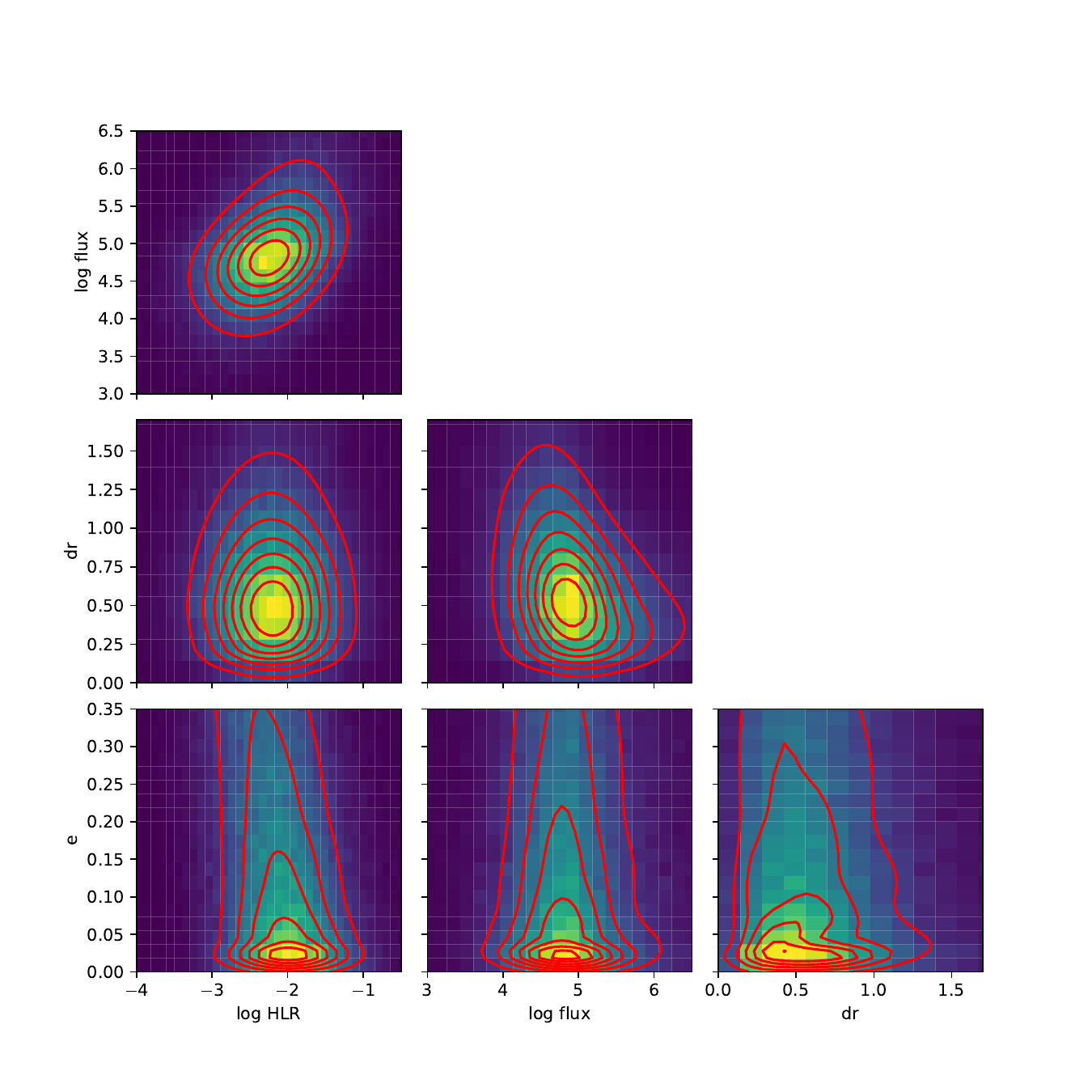}
\caption{Two-dimensional marginal densities of the BGMM model for disks. Yellow and blue indicate histogram bins of higher and lower density, respectively.\label{fig:discs_2d}}
\end{figure}

\begin{figure}
\includegraphics[width=\textwidth]{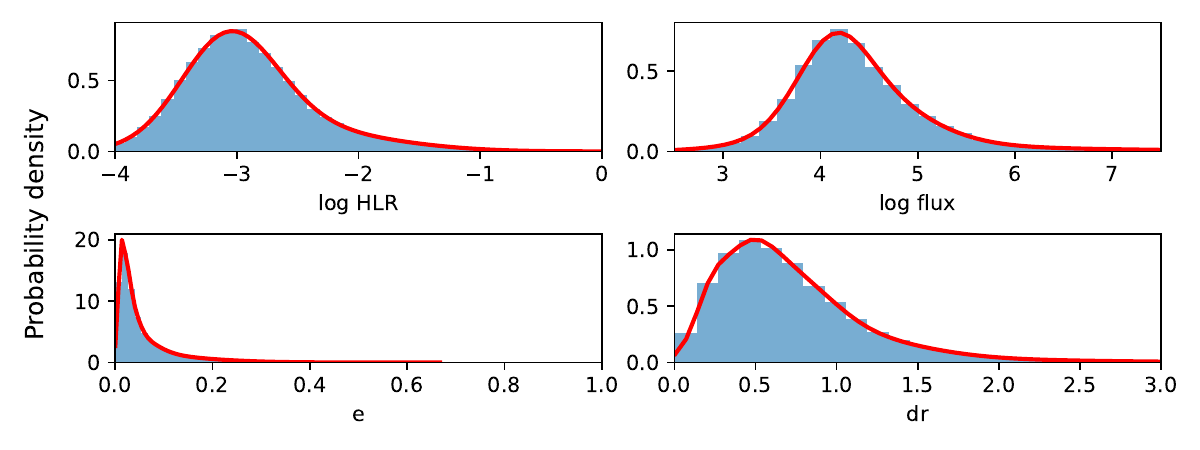}
\caption{1-dimensional marginal densities of the BGMM model for bulges.\label{fig:bulges_1d}}
\end{figure}

\begin{figure}
\includegraphics[width=\textwidth]{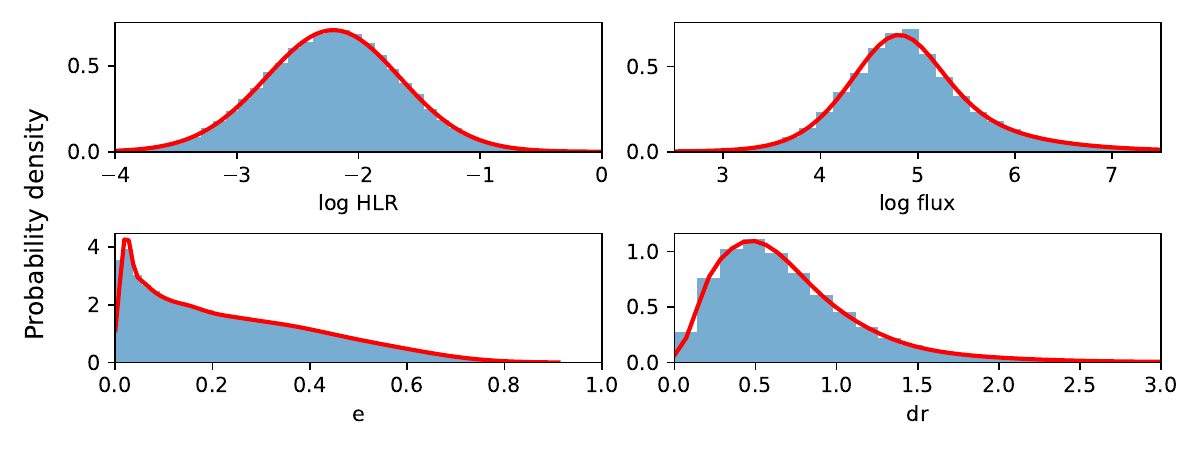}
\caption{1-dimensional marginal densities of the BGMM model for disks.\label{fig:discs_1d}}
\end{figure}

Because these distributions strongly correlate with galaxy morphology in the CosmoDC2 catalog, we perform separate fits for bulges and disks. In each case,\footnote{Aside from the numbers of components (together with our procedure for finding them) and the number of iterations needed for convergence (which went as high as 6000), what follows are all just the default settings for scikit-learn's \texttt{BayesianGaussianMixture} class as of the time of writing.} an idealized infinite-component, Dirichlet process mixture model is approximated using a Stick-breaking representation with a finite maximum number of components. The number of components in each fit, 36 for bulges and 28 for disks, was chosen by looking for the largest number that produced a pareto improvement of both the likelihood and the evidence lower bound (ELBO) score for each model, given a ``patience'' of 5 additional components to seek further improvement. Responsibilities for each component were initialized using the kmeans algorithm. All fits were run to convergence, defined as an expectation-maximization iteration with a lower-bound average gain on the likelihood below 1e-3. Each Gaussian component has its own, general covariance matrix. A small regularization value of 1e-6 was added to the diagonal elements of each covariance matrix to ensure they would all be positive at every computational step. Figures~\ref{fig:bulges_2d} through \ref{fig:discs_1d} show various marginal projections of the full four-dimensional BGMMs (the contours and density curves plotted in red), compared to the true data distributions (the histograms).


\subsection{Likelihood}\label{sec:likelihood}
By likelihood, we mean the probability density for a given galaxy light profile to produce a specific array of footprint pixel values.\footnote{For a fixed footprint pixel array shape. Probabilistically modeling the footprint shape would involve resimulating the detection process at each step of the MCMC chain, which we do not do here.} By a galaxy's light profile, we mean the time-averaged distribution of galactic light hitting the top of Earth's atmosphere. The same light profile will randomly produce different images due to factors such as sky noise and photon shot noise. We model the likelihood by assuming that the intensity of each coadd image pixel is Gaussian-distributed and statistically independent of all other pixels. For a given set of parameters describing the light profile and PSF, we take the mean of each pixel's likelihood to be the convolution of the light profile with the PSF, integrated over that pixel. This is equivalent to simulating a noiseless image of the galaxy after convolving it by the PSF, all of which we do using \texttt{GalSim}.

The Pipelines estimate a level of variance in each coadd pixel, consistent with astrophysical shot noise and assumed sky noise. By assuming shot noise is Poisson, the Pipelines estimate that variance increases roughly linearly with image pixel brightness. When we compare the instrumental flux reported in each coadd pixel to the Pipelines-estimated variance, we see that this linear relationship tends to hold well, albeit somewhat noisily. In each patch, we therefore perform a linear fit of instrumental flux vs.\ variance on all pixels in the patch, except for ones flagged with any of the pathologies noted in Section~\ref{sec:footprints}. The amount of shot noise we assume in each pixel of a simulated image is the slope of this fit times the simulated image's instrumental flux; the amount of background noise we assume in each pixel is the fit intercept. The lead author of the present study originally developed this method for injecting semi-realistic variance into LSST coadds for the purpose of SSI, code for which is included in the Pipelines.\footnote{\href{https://github.com/lsst/source_injection}{\nolinkurl{github.com/lsst/source_injection}}} More broadly, the code used for simulated image generation in this study is closely equivalent to the relevant Pipelines SSI code.

\section{Selection corrections}\label{sec:selection_corrections}
\subsection{Realistic and Simplified Datasets}\label{sec:data_sets}
Even though we have taken steps to construct an idealized model corresponding in many ways to the true prior distribution of galaxy parameters and to the true properties of the data-generating processes (i.e., the likelihood), there remain significant differences between our assumed probabilistic model and these true processes. In reality, even for the DC2 dataset, the PSF is generally more complex than the simple Kolmogorov form, and it can change discontinuously within an image for coadds constructed via the present Pipelines. Our assumed variance comes from a linear fit with a relationship that is assumed to be constant within images, so for coadds, this will not exactly match the variance of the true data-generating process. The same is true for our photometric calibrations. The single-component Sérsic profiles we assume for each galaxy do not exactly correspond to the bulge-plus-disk-plus-knots profiles that were actually used in the DC2 image simulations. Coaddition introduces covariance between pixels that we have completely ignored in our likelihood model. Electronic artifacts such as the brighter--fatter effect \citep{2015AandA.575.A41} and other technical features of real data also lack a place in our current model.

One therefore anticipates some level of mismatch, on average, between the truth and the estimated posteriors under our model. To see the specific impact of these simplifying modeling assumptions, we generated a second, ``simplified'' dataset. This has the same distribution of \textit{true} galaxy properties as the DC2 data, but here, we construct \textit{images} of those galaxies in \texttt{GalSim} in a manner that precisely matches our modeling assumptions, including the form of the PSF, the assumed variance plane, the photometric calibrations, and a pure bulge or disk morphology for every galaxy. By way of contrast, we refer to the DC2 images as the ``realistic'' dataset. We then ran essentially the same MCMC fitting procedure on both datasets; the following section describes the one difference.

\subsection{Conditioning on Selection}\label{sec:conditioning_on_selection}
The Pipelines detection algorithm defines a $5\sigma$ detection threshold for the flux of an object with a point-source morphology. Sometimes an object with a true mean flux right at the $5\sigma$ level will produce enough light to be detected, but just as often it will not. The footprints in the \textit{realistic} dataset, therefore, only include images of $5\sigma$ objects containing more than the average amount of flux such objects are expected to produce. If no correction is applied, this significantly skews the likelihood and biases the resulting posteriors.

More specifically, the realistic footprints must pass three significant criteria to be selected for analysis. First, the pixel values must be such that they pass the Pipelines detection algorithm, which broadly means that the observed instrumental flux must be some combination of very bright and sufficiently PSF-like. Second, among the true objects that the footprint pixels overlap, one of those true objects must have a flux at least 50 times larger than the total flux of all the others, and none can be stars or supernovae. Finally, the footprint must have exactly one detection peak.

These criteria mean that the likelihood for the realistic data must be conditioned on the fact of selection. Formally, for any given data example,
\begin{equation}
p(\theta | D, det) = p(\theta | det) p(D | \theta, det) / p(D | det),
\end{equation}
where $\theta$ is the 9D parameter vector describing the true properties of the galaxy and PSF, $D$ is the observed array of image pixel values, and $det$ is the condition that this data example has passed our selection criteria. Since $det$ directly determines the set of true $\theta$ values considered in this study, the conditional prior $p(\theta | det)$ ought to reflect that fact for \textit{both} the simplified and the realistic datasets. This is precisely the prior described in Section~\ref{sec:prior}.

Once $\theta$ is specified, the condition $det$ does not influence $D$ in the \textit{simplified} dataset, because any rendered $D$ is analyzed for every example. The conditional likelihood $p(D | \theta, det)$, therefore, reduces to $p(D | \theta)$, which is just the Gaussian likelihood we described in Section~\ref{sec:likelihood}. However, in the realistic dataset, the fact of detection $det$ deeply influences $D$.

By making the approximations described in Appendix~\ref{sec:detection_correction_appendix}, we can write the selection-conditioned likelihood as
\begin{equation}
p(D | \theta, det) = p(D | \theta) / \int_{\mathcal{D}et} dD' p(D' | \theta).
\end{equation}
The denominator is the total probability that a galaxy with true parameters $\theta$ would produce any observed image pixel array $D'$ that would be detected by the Pipelines. We can place an approximate lower bound on this integral by computing the probability that the pixel in the detection image (the original image convolved with a Gaussian approximation of the PSF) covering the galaxy's center would be bright enough to pass the $5\sigma$ threshold.

Since the likelihood is assumed to be Gaussian and independent in every image pixel, the detection image has a Gaussian distribution in the center-covering pixel as well. We compute its mean by taking the galaxy image we use for likelihood calculations and convolving it with the same approximate PSF used by the Pipelines for detection. Its variance is the value of the center-covering pixel of the variance plane convolved with the square of the PSF. The integral above goes from the minimal detection threshold (which we take from the Pipelines) to infinity, which we compute using \texttt{scipy.stats}.

Since the integral $\int_{\mathcal{D}et} dD' p(D' | \theta)$ is the probability that any image produced by $\theta$ would pass the Pipelines detection algorithm but not necessarily any other selection criteria, another lower bound on this integral is the probability that any image produced by $\theta$ would pass all of our selection criteria. We estimate this probability empirically by finding, within various bins of flux, the fraction of galaxies that ultimately give rise to a footprint passing all of the selection criteria. Our final estimate for $\int_{\mathcal{D}et} dD' p(D' | \theta)$ is the maximum of the two lower-bound estimates described above.

\section{MCMC}\label{sec:mcmc}
\subsection{Initializing and Running the Chains}\label{sec:running}
The functions for computing the prior and likelihood are implemented in \texttt{JIF}, an open-source Python framework for organizing these computations for astronomical images and running MCMC. Given the above prior and likelihood model, we use \texttt{JIF} to estimate the corresponding posterior for each footprint with an affine-invariant ensemble MCMC method described in \citet{2010CAMCS.5.1} (hereafter G\&W), as implemented in \texttt{emcee} \citep{2013PASP.125}. This has the practical advantage of requiring relatively little parameter tuning. The chains consist of sequences of 9D vectors that model the nine parameters of interest for the galaxy and PSF in a given footprint.

Rather than a single sequence for each footprint, a chain in the G\&W method comprises some number of parallel ``walkers.'' We set the number of walkers to 18, because \texttt{emcee} cannot fit $n$ parameters with fewer than $2n$ walkers, and in preliminary experiments, we did not find any advantage from using more. The chains progress via ``walk moves'' as defined by G\&W, using the full walker complement to estimate the covariance matrix for each proposal.

All MCMC runs reported in this study consist of 1000 burn-in steps followed by 3000 posterior inference steps. In initial tests with a handful of diverse footprints, these numbers of steps consistently yielded highly stable estimates of the posterior that don't significantly change when the chain is re-run or when the number of steps is increased. These features suggest robust convergence, which is also indicated by the results shown in Section~\ref{sec:results}.

To avoid a lengthier burn-in phase as well as potentially pathological (and highly unrealistic) regions of parameter space for which \texttt{GalSim} cannot render images, we start each chain at parameter values near a MAP estimate of those parameters. We use the L-BFGS-B algorithm \citep{1995SIAMJSC.16.5} as implemented in \texttt{scipy.optimize.minimize} to minimize the negative-log-(likelihood times prior), at which point one has the MAP. Occasionally, L-BFGS-B fails, in which event we retry the fit with the SLSQP algorithm \citep{SLSQP}; one or the other of these nearly always succeeds. Each of the 18 walkers starts at an independently sampled point in parameter space, with each parameter independently sampled from a Gaussian distribution, having a mean equal to the estimated MAP value and a standard deviation of one percent of that value.

To initialize the initial MAP fit, both ellipticity components are set to 0, and the centroid position is set to the center of the bounding box. The initial flux is estimated by taking the sum of the footprint image pixels and inverting the photometric calibrations to obtain a total light profile flux in nJy, assuming a footprint aperture correction of 1.0 (see Appendix~\ref{sec:photometric}). The initial half-light radius is then set to its expected value conditioned on the aforementioned flux estimate, under a greatly simplified ``interim prior'' consisting of a 2D Gaussian fit of the true distributions of log-flux and log-HLR of galaxies in our dataset (using separate fits for bulges and disks, as with our full prior). In instances of failed MAP fits, we initialize the chains using these rougher estimates, which are often rather similar to the MAP values.

Each of these steps---a first rough initialization, followed by an MAP fit, followed by MCMC---requires increasing amounts of time but offers increasing accuracy and insight. The first step can be done almost instantly, but its estimates have only a loose relationship to the posterior we are trying to estimate. It might perform just as well as an MAP fit to initialize the MCMC chains (1000 burn-in steps does a good job of getting a chain to converge starting from just about anywhere), but we settled on the MAP because it does not extend the total compute time by a significant fraction, while, by definition, being in the part of parameter space that a well-converged MCMC chain should sample from most often. The MAP fit generally requires, at most, a few hundred simulated likelihood and prior computations per footprint.

For this study, we fit 116,088 total footprints, requiring more than 8000 CPU hours for the MCMC alone, rendering 8.4 billion simulated galaxy images in the process. The mean time per model fit was 4.2 minutes. Because we fit each footprint independently, the MCMC run for each footprint can be computed in parallel. Using 2744 cores at Lawrence Livermore National Laboratory (LLNL), the entire suite of MCMC computations was completed in about three hours of wall clock time. The preceding numbers apply to the simplified dataset on its own; the realistic dataset required similar compute.

Our dataset includes every isolated $i$-band footprint that lies inside both DC2 tract 3830 and one of CosmoDC2 HEALPixels 9685, 9813, or 9814, which together comprise a contiguous region of space covering most of that tract. Of the 232,176 corresponding chains we analyze (one each for the realistic and simplified versions of every footprint), all but 178 ran to completion without errors. The success rate is not always this high for all possible priors.

\subsection{Informed vs.\ Uniform Priors}\label{sec:uniformprior}
As noted earlier, most Bayesian galaxy fitting studies use a uniform prior over the fit parameters. This has the predictable consequence of ``overfitting'' low-S/N images. Sampled parameter values can exceed hard boundaries, which are either set explicitly or emerge implicitly from the practical inability to compute likelihoods with certain parameter combinations.

This seems to be the underlying source of the large number of fit failures reported in the \texttt{ProFit} analysis by \citet{2017MNRAS.466.2}. As is typical among studies introducing Bayesian galaxy modeling codes, \citet{2017MNRAS.466.2} described their likelihood model in detail, but presumably used an implicit uniform prior. Such a prior was in fact justified in that study, because their simulated dataset was generated from a uniform distribution of model parameters. Nevertheless, they observed an increasing number of fit failures as galaxy S/N decreased, tallying up a proportionally rather large number of fit failures in total.

\begin{figure}
\includegraphics[width=\textwidth]{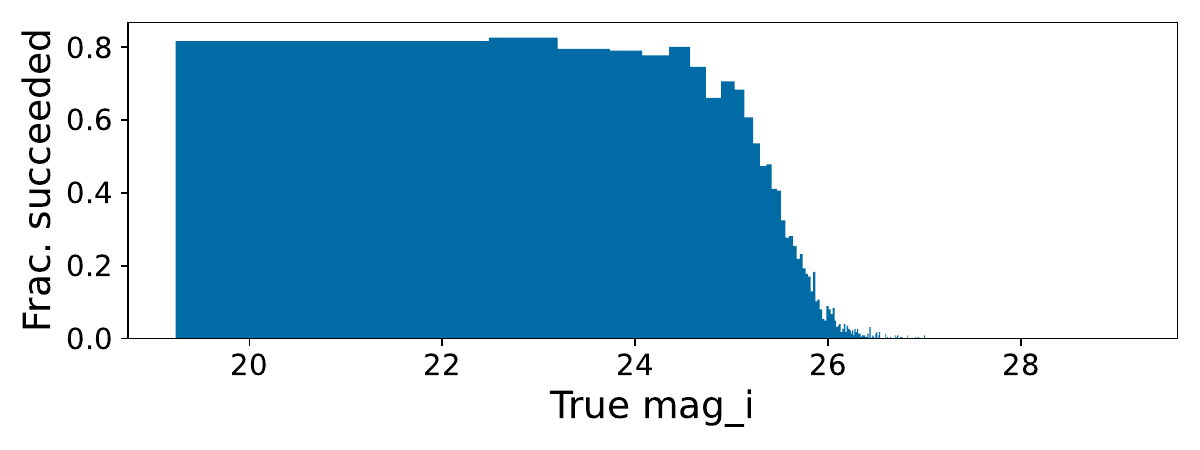}
\caption{MCMC success rate with a uniform prior. Each bin contains the same number of galaxies. The equivalent plot with our informed prior is just a solid blue rectangle with a height of 1.0. Simplified data, bulges only.\label{fig:fitfailures}}
\end{figure}

We reproduce the same phenomenon here by rerunning our fitting procedure with a uniform prior. The only boundaries we explicitly place on the fit parameters are absolute mathematical constraints such as the fact that half-light radius must be positive. As shown in Figure~\ref{fig:fitfailures}, we observe an increasing number of fit failures as S/N decreases, with the success rate settling around 82\% for high-S/N footprints and 0\% for low-S/N. These failures stem from failed attempts to render model images from sampled parameters. Meanwhile, even for successful fits, the time required to render images can explode when the parameters are too extreme. To conserve CPU hours, for the uniform-prior fits, we cap the number of chain steps at 2000 instead of 4000, and we give each chain a maximum of 10 minutes to run.

The PSF arises due to well-understood effects that take place ``close to home'' and that are, to a certain extent, measured directly during data taking; so even for these fits, we keep an informed prior on the distribution of PSF parameters. In practice, we observe that this helps prevent pathological fits in which the estimated PSF width becomes extremely unrealistic, probably spurred by its degeneracy (in terms of the resulting image) with galactic width.

\section{Results}\label{sec:results}
When interpreting the plots below, recall that we fit the true bulges (more precisely, the very bulge-like galaxies) with a pure bulge light profile, and use a prior that has been fit to the true distribution of bulge-like galaxy properties in our dataset; disks are fit with a corresponding disk light profile and prior. This means that all results reported in this section apply only to a morphologically preselected 81\% subsample, with performance on the minority of more intermediate cases left untested. For now, we leave fitting alternative possibilities for the galaxy morphology to other stages of hierarchical inference not developed here.

Because it is the one we understand best, we mostly show results for the simplified dataset, but we show enough from the realistic data to indicate the effect of selection corrections as well as some key points of needed development. For brevity's sake we mostly just show bulge distributions.

\subsection{Chains}
\begin{figure}
\includegraphics[width=\textwidth]{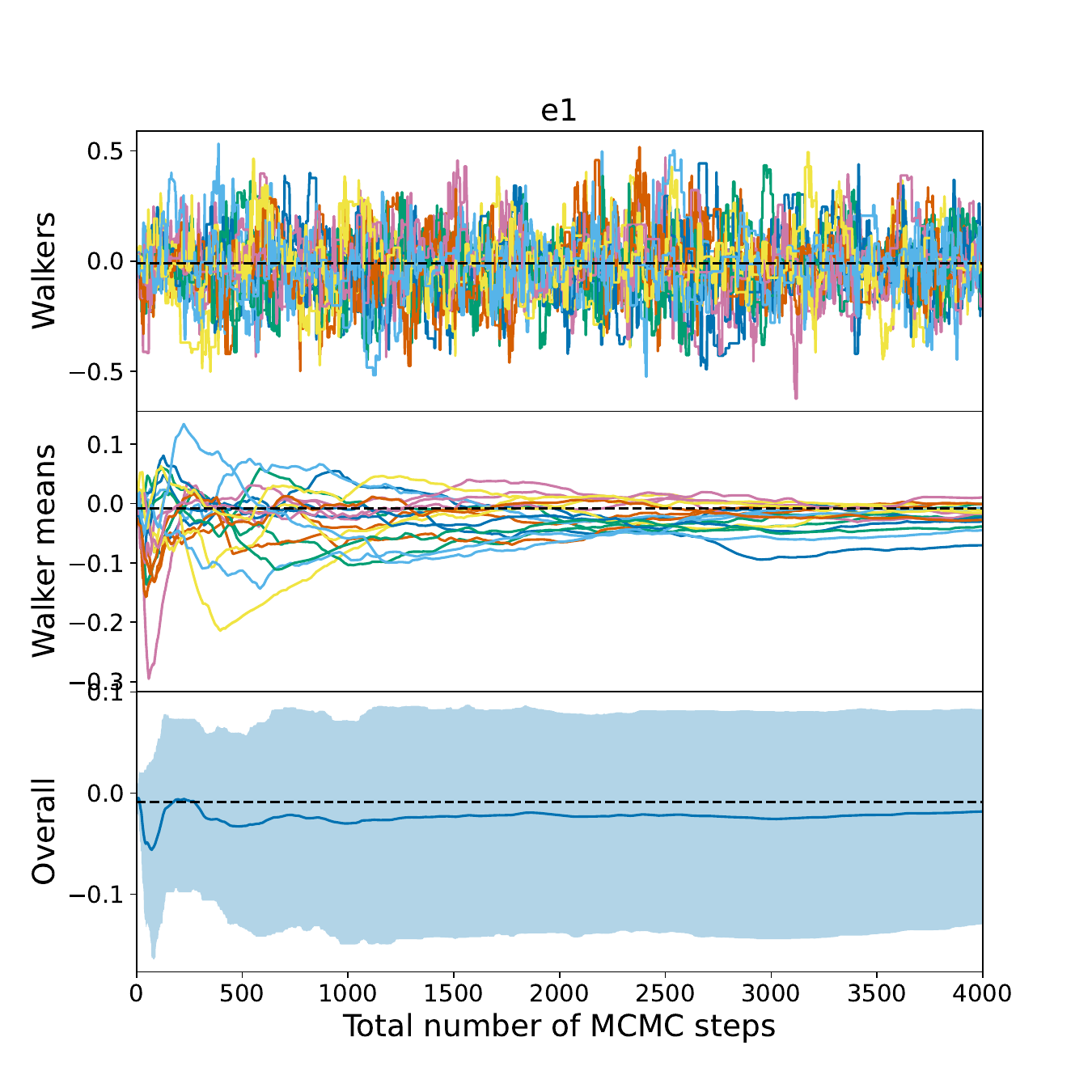}
\caption{Illustrations of MCMC for a single galaxy in the simplified bulge-only dataset with mag 26, focusing on e1. Top: e1 for each walker, over each of the 4000 MCMC steps. Middle: walker means up to each step. Bottom: mean of all e1 values across all walkers up to each step (solid line), together with the spread from the 16th--84th percentile. The true e1 is indicated in all plots by the dashed line. Means and spreads at each step $n$ ignore the first $n/4$ steps.\label{fig:trace_hlr}}
\end{figure}

Figure~\ref{fig:trace_hlr} shows several views on the progress of the e1 values of a single chain, consisting of 18 parallel walkers iterating for 4000 steps each. The top plot shows the trace of each separate walker's e1 value at every step. The individual traces exhibit so much random variability that it can be difficult to discern meaningful trends.

The middle plot shows the mean of each walker's complete trace up to each step, ignoring the first quarter of steps (up to each point) as burn in. After a period of initial volatility, the walker means settle into steady behavior over the course of several hundred steps, smoothing out over several thousand steps. The bottom plot shows the overall mean of the combined values of all walkers, together with the spread from the 16th--84th percentile of all walker values, up to each step (ignoring the first quarter of steps up to each point).

\begin{figure}
\includegraphics[width=\textwidth]{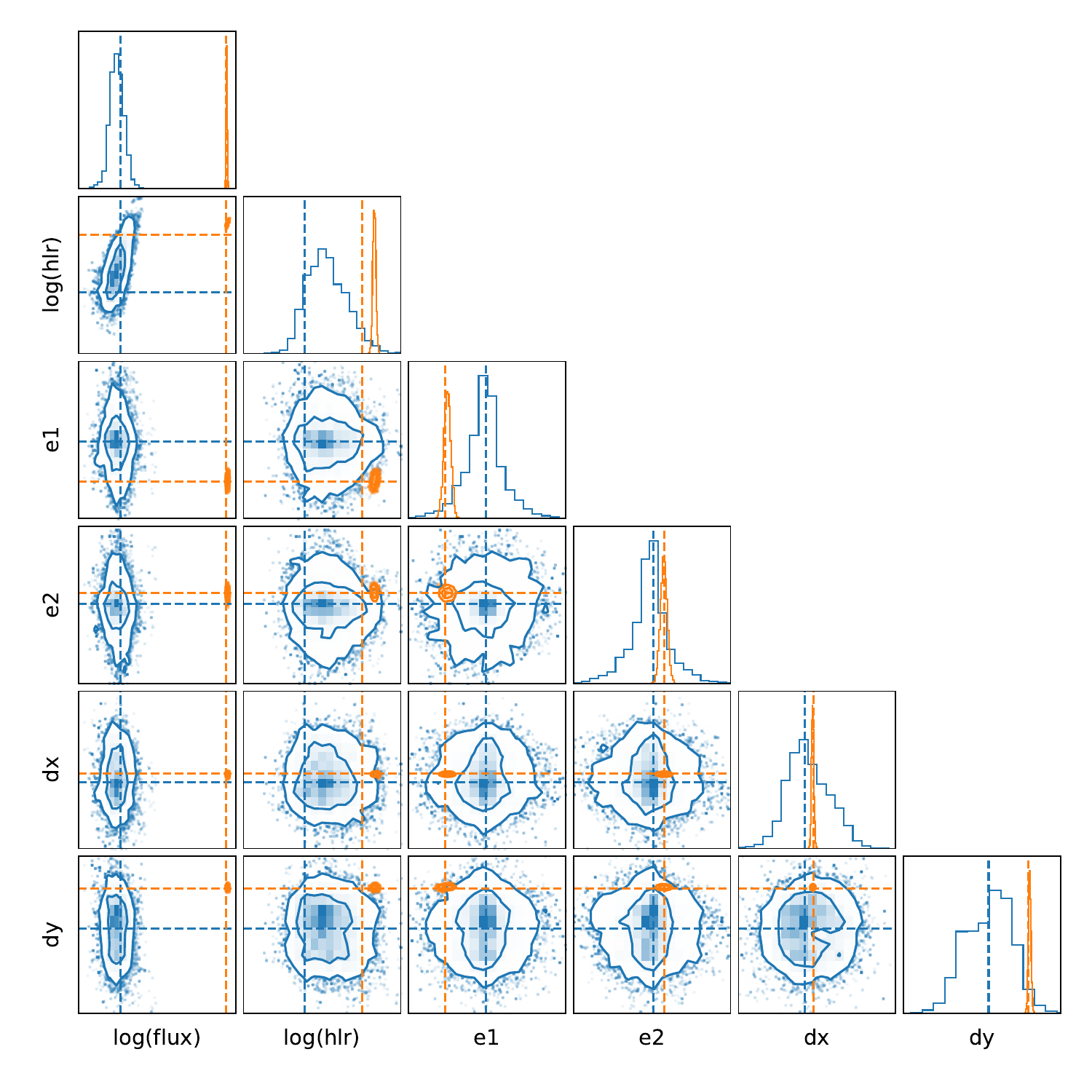}
\caption{MCMC-estimated posteriors for two different examples from the simplified bulge-only dataset. Blue corresponds to an example with mag 26, near the detection threshold. Orange is an example with mag 22. Inner contours of the 2D plots (when visible) show 68\% credible regions, while outer contours show 95\% regions. Dashed lines show truth.\label{fig:cornerplot}}
\end{figure}

Figure~\ref{fig:cornerplot} illustrates the 6D posteriors on the galaxy parameters for two different examples: one with a mean flux close to the detection threshold, for which the prior largely determines the posterior, and one with a flux bright enough to be predominantly likelihood-constrained. For this and all other figures showing final posterior estimates, burn-in steps have been removed. As expected, the posterior widths decrease as the true brightness increases. The relative amount of decrease depends on the parameter: the dx--dy contours shrink almost to a point, whereas the e1--e2 contours remain clearly spread out, indicating that e1 and e2 require more signal to infer compared to dx and dy.

The true parameter values for each example lie inside the outermost relevant 2D contours in most cases and often the inner contour as well. We explore the posterior calibration more systematically in Section~\ref{sec:truth_comparison}. In this plot, we have omitted the PSF values for visual clarity, primarily because they don't show much: even at mag 22, there is not enough information to noticeably affect the PSF posterior. Only the very brightest examples in our datasets are able to shrink the PSF posterior widths by a perceptible amount. All of the key results in this paper continue to hold if we omit PSF modeling entirely, but we have included it as a demonstration that the PSF can be jointly included in Bayesian galaxy modeling, and also that our fitting method can straightforwardly accommodate nine parameters.

\subsection{Chain Properties}\label{subsec:chain_properties}
\begin{figure}
\gridline{\fig{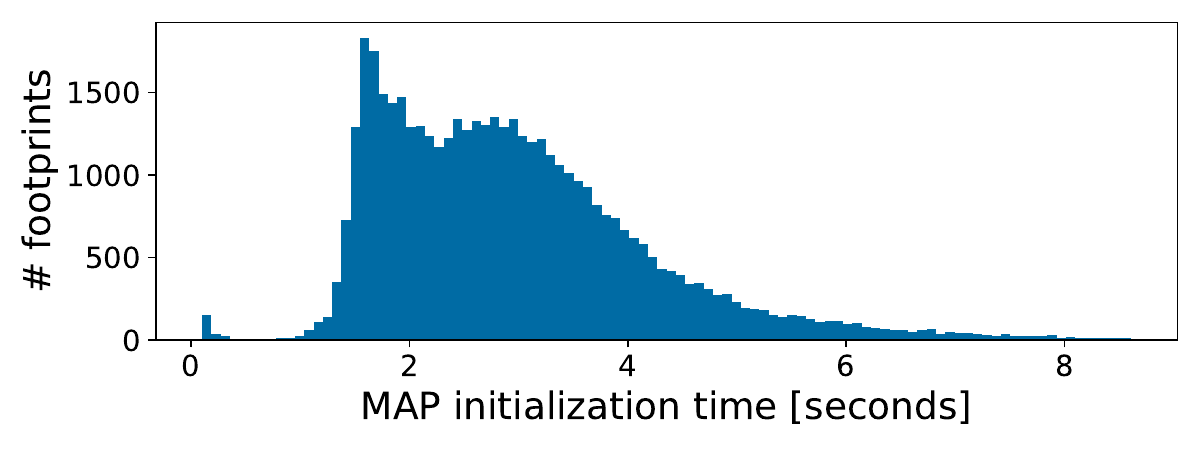}{0.5\textwidth}{(a)}
          \fig{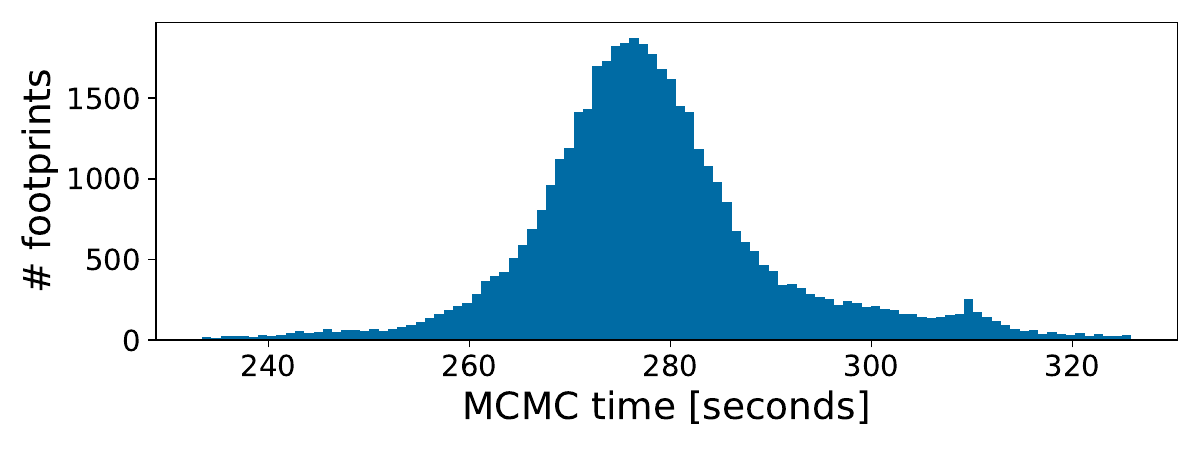}{0.5\textwidth}{(b)}}
\gridline{\fig{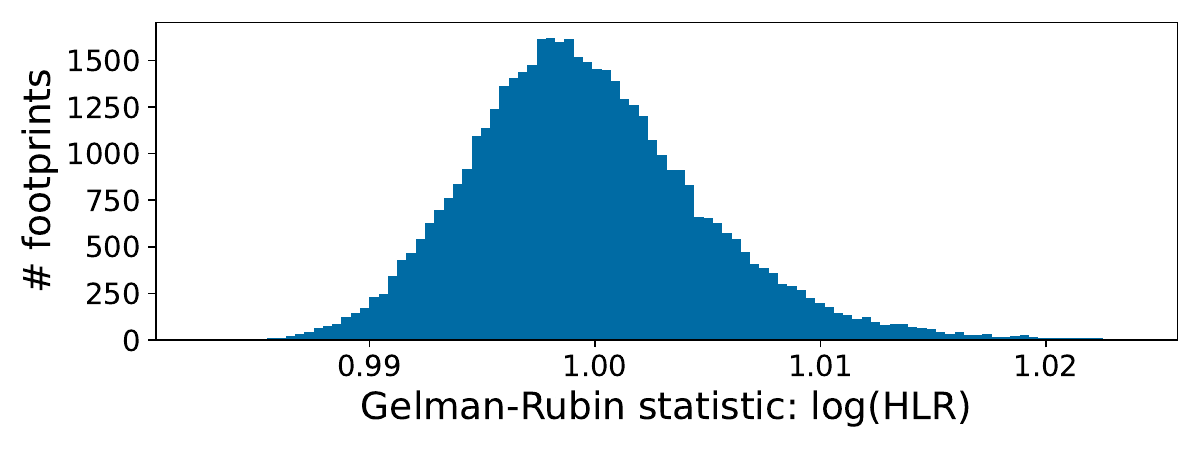}{0.5\textwidth}{(c)}
          \fig{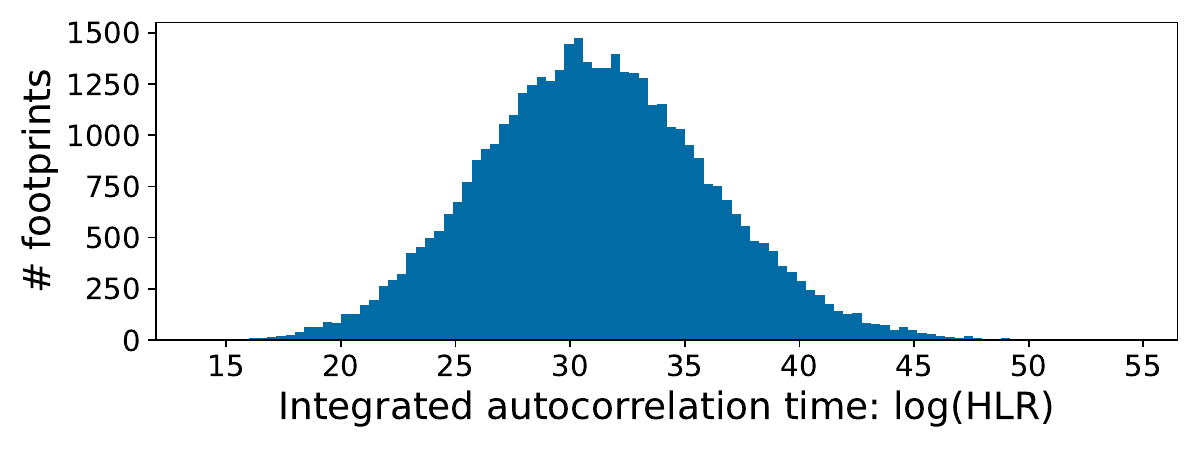}{0.5\textwidth}{(d)}}
\caption{(a),(b): Distribution of times needed to (a) initialize the chains and then (b) run 4000 steps of MCMC. (c),(d): Distribution of (c) Gelman--Rubin statistic values and (d) integrated autocorrelation times. Simplified dataset, bulge galaxies only.\label{fig:times_and_chainstats}}
\end{figure}

Figures~\ref{fig:times_and_chainstats}(a) and (b) show the distributions of times needed for chain inference in the simplified dataset of bulge galaxies (any distributions noted in this section are largely similar for all data variations). Initial MAP estimation takes between 1 and 2 seconds on average, while MCMC takes 4 minutes on average. These run times are similar on a modern commodity laptop and on one core of a high-performance computing node at LLNL.

These run times compare favorably with \texttt{pysersic}, which takes ``a few minutes or less on modern laptops'' to fit a seven-parameter Sérsic profile \citep{2023JOSS.pysersic}; \texttt{AstroPhot}, which took 738.5 seconds to sample a 13-parameter model of various components in a scene \citep{2023MNRAS.AstroPhot}; and \texttt{BANG}, which needed about 30 minutes per galaxy to fit three different morphological models with 14--18 parameters each \citep{2023MNRAS.525.1}. We stress that none of these comparisons are absolute, because each analysis used a distinct computer architecture, modeled a distinct parameter set, and made a distinct determination of the number of MCMC samples necessary for statistical confidence in the results. Nevertheless, it is noteworthy that the single-model run times for all of these approaches ended up in the same ballpark.

Figure~\ref{fig:times_and_chainstats}(c) shows the distribution of Gelman--Rubin~\citep{1992StatSci.7.4} values for log-half-light radius (HLR) in simplified bulge footprints. Most values lie between 0.99 and 1.01; a value of 1.0 is generally considered to indicate high-quality convergence. Each parameter has its own value for this statistic, though we find that the distributions for every parameter look quite similar. The interpretation of the Gelman--Rubin statistic is muddied when using correlated parallel chains, so as an alternative, \citet{2013PASP.125} defined a measure of integrated autocorrelation time applicable to such chains. Figure~\ref{fig:times_and_chainstats}(d) shows the distribution of integrated autocorrelation times for log-HLR. Again, the distributions of this value for all parameters are all similar. Because we use 3000 MCMC steps for posterior estimation after burn-in, and the integrated autocorrelation time is generally between 20 and 40 steps, we conclude that more than 75 fully independent posterior samples were drawn for each parameter in nearly every footprint.

\subsection{Comparison to Truth}\label{sec:truth_comparison}
\begin{figure}
\gridline{\fig{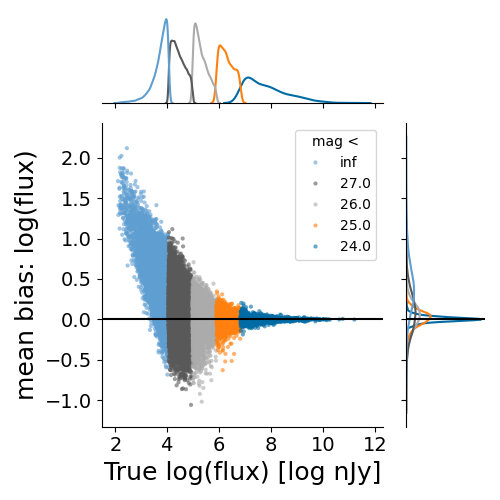}{0.5\textwidth}{(a)}
          \fig{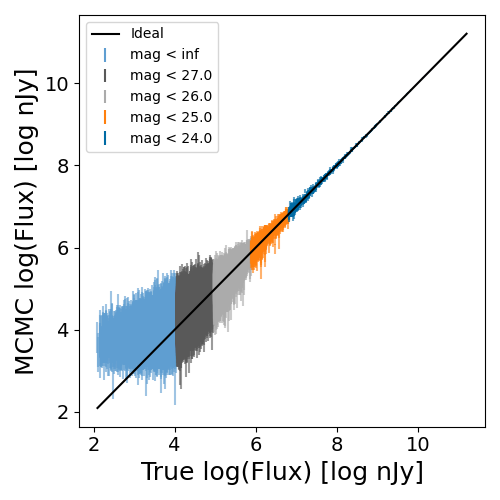}{0.5\textwidth}{(b)}}
\caption{(a): Difference between the MCMC estimate of the posterior mean and truth for each galaxy, for the natural logarithm of the total flux. The marginal contours show the overall distribution of points in each mag bin. (b): The vertical lines show the interval from the 16th--84th percentile of the distribution of MCMC values, plotted vs.\ truth on the x-axis. Simplified data, bulges only.\label{fig:bias_logflux_simplified_bulge}}
\end{figure}

Figure~\ref{fig:bias_logflux_simplified_bulge}(a) shows the difference between the posterior mean and the true value of the natural logarithm of the total flux. The marginal curves on top are normalized to highlight their separate shapes. As shown by the well-centered marginal curves on the right, galaxies brighter than mag 26 have virtually unbiased posterior means, when averaged over many different galaxies. As expected, the spread of values decreases as the sources get brighter.

\begin{figure}
\gridline{\fig{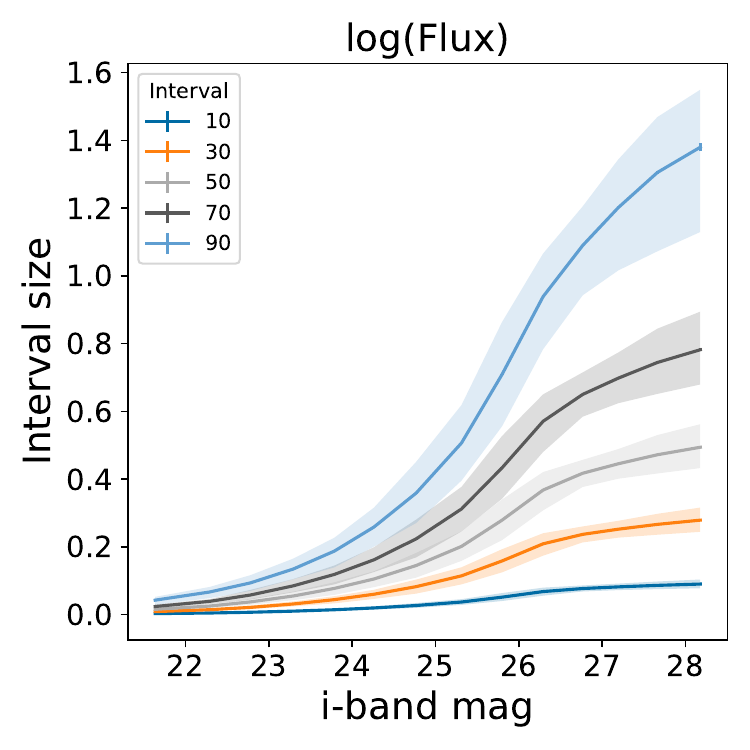}{0.5\textwidth}{(a)}
          \fig{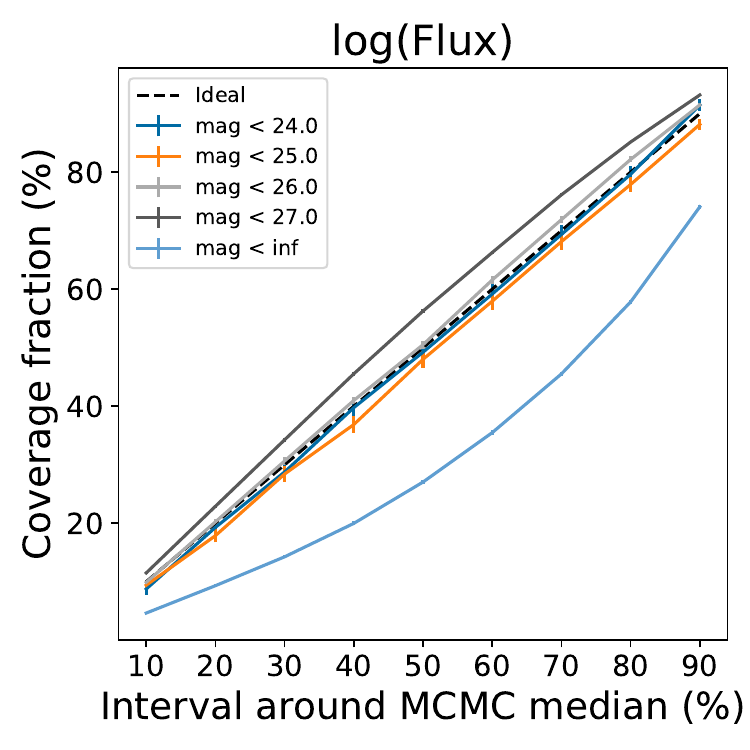}{0.5\textwidth}{(b)}}
\caption{(a): MCMC-estimated credible intervals, plotted vs.\ true galaxy brightness. The lines are the mean credible interval size (for the interval noted in the legend) for galaxies of a given flux level, while the bands show the 68\% spread in the widths among the examples we fit. Statistical uncertainty due to finite dataset size is too small to appear on this plot. (b): Fraction, on the y-axis, of instances in which an interval of a given size, on the x-axis, overlaps the truth. The vertical error bars indicate statistical uncertainty due to finite dataset size. Simplified data, bulges only.\label{fig:intervals_logflux_simplified_bulge}}
\end{figure}

Figure~\ref{fig:bias_logflux_simplified_bulge}(b) compares the posterior directly to the truth, this time showing the 68\% inner credible interval for each posterior. Just as the posterior means adhere more closely to the truth as the true galaxy brightness increases, so too do the credible intervals shrink. This tendency is illustrated comprehensively in Figure~\ref{fig:intervals_logflux_simplified_bulge}(a), which shows how the credible intervals of any given percentage width shrink as the brightness of a galaxy increases.

Credible intervals at all brightness levels generally overlap the truth the expected fraction of the time, as illustrated in Figure~\ref{fig:intervals_logflux_simplified_bulge}(b). If the estimated posteriors are reliable, then an $n$-percent credible interval should overlap the truth $n$ percent of the time, on average over many galaxies, resulting in a perfectly diagonal plot. We see that this broadly holds to within statistical errors for much of our data, except for the dimmest galaxies. The direct cause of this discrepancy is clear from either of the two preceding figures: the MCMC-estimated log-flux is systematically higher than the true flux for these dim galaxies.

More specifically, as shown in Figure~\ref{fig:bias_logflux_simplified_bulge}(b), the estimated posterior for log-flux is basically constant for all galaxies dimmer than mag 27. This is because the observed pixel data for such galaxies provides essentially no information about the underlying galaxy. The minimum detection threshold is brighter than mag 26 in every patch, meaning that the dimmest galaxies only enter our dataset if their DC2 images coincided with random noise fluctuations that just happened to rise above the detection threshold. Once selection corrections have been taken into account, the likelihood of observing any given footprint image in our realistic dataset is essentially independent of the underlying galaxy brightness for galaxies dimmer than mag 27.
\begin{figure}
\gridline{\fig{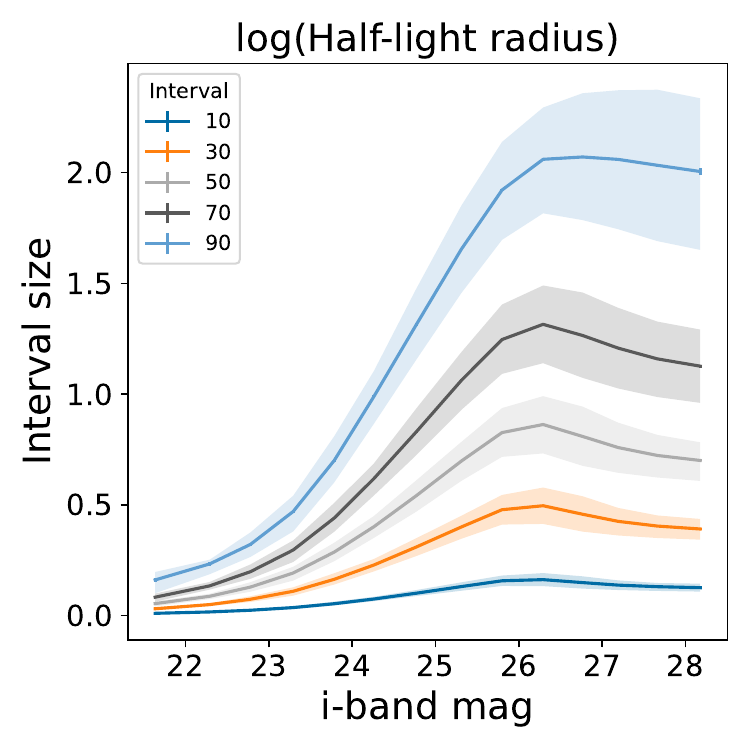}{0.5\textwidth}{(a)}
          \fig{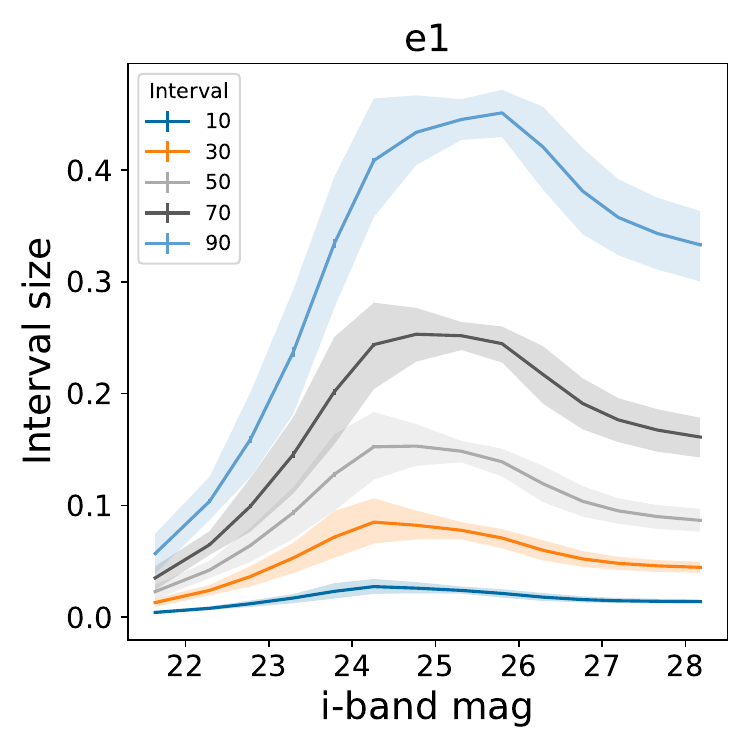}{0.5\textwidth}{(b)}}
\gridline{\fig{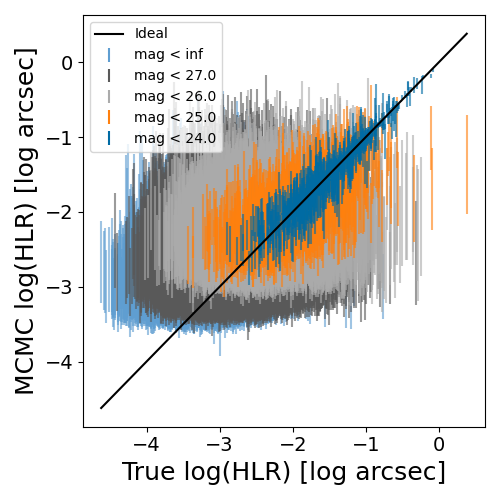}{0.5\textwidth}{(c)}
          \fig{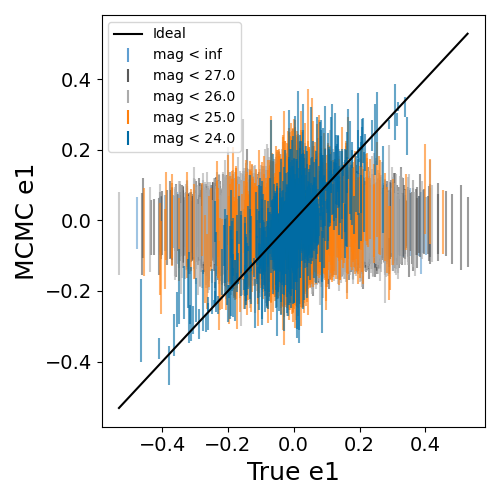}{0.5\textwidth}{(d)}}
\caption{(a),(b): The widths of various MCMC-estimated intervals of the posterior, plotted vs.\ true galaxy brightness. (c),(d): Inner 68\% posterior intervals, plotted vs.\ truth. Simplified data, bulges only.\label{fig:hlr_e1_simplified_bulge}}
\end{figure}

\begin{figure}
\gridline{\fig{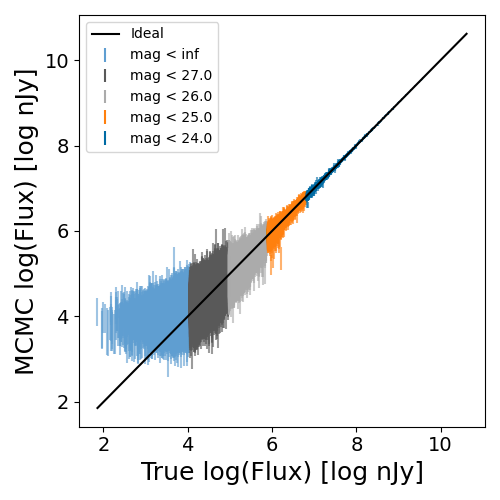}{0.5\textwidth}{(a)}
          \fig{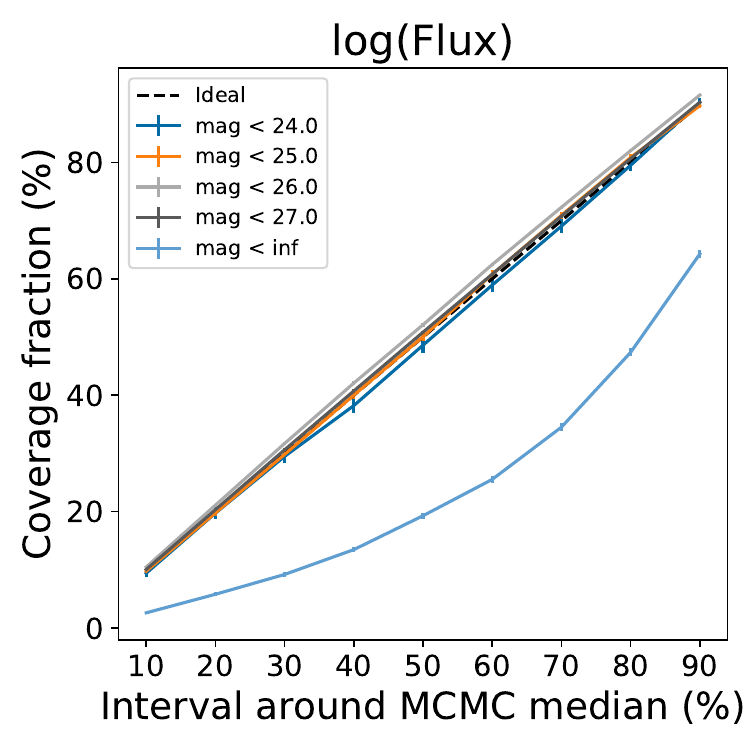}{0.5\textwidth}{(b)}}
\caption{(a): Inner 68\% posterior intervals for log-flux, plotted vs.\ truth. (b): Posterior calibration for log-flux. Simplified data, disks only.\label{fig:logflux_simplified_disc}}
\end{figure}

Since the posterior is proportional to the likelihood times the prior, and since the shape of the likelihood is highly consistent for any data likely to be produced from the dimmest galaxies, the posterior becomes largely independent of the data for such galaxies. The flux level at which this occurs in our data is illustrated by the flat section of the curves on the right side of Figure~\ref{fig:intervals_logflux_simplified_bulge}(a), as well as by the flattening of the left section of Figure~\ref{fig:bias_logflux_simplified_bulge}(b), which, in turn, translates to the diagonal ``swoosh'' on the left section of Figure~\ref{fig:bias_logflux_simplified_bulge}(a).

The fact that the constant posteriors are shifted upward from the truth in the dim-galaxy regime corresponds to the fact that our prior is strongly peaked at mag 26, which is consistent with how these posteriors accumulate below mag 26 ($\text{log-flux} < 5$) in Figure~\ref{fig:bias_logflux_simplified_bulge}(b). Since the posterior for almost every galaxy in the dimmest flux regime has almost exactly the same too-high mean and the same width, irrespective of the true galaxy brightness, credible intervals of any size consistently sit above where they need to be, resulting in the pronounced undercoverage illustrated by the light-blue curve in Figure~\ref{fig:intervals_logflux_simplified_bulge}(b).
\begin{figure}
\gridline{\fig{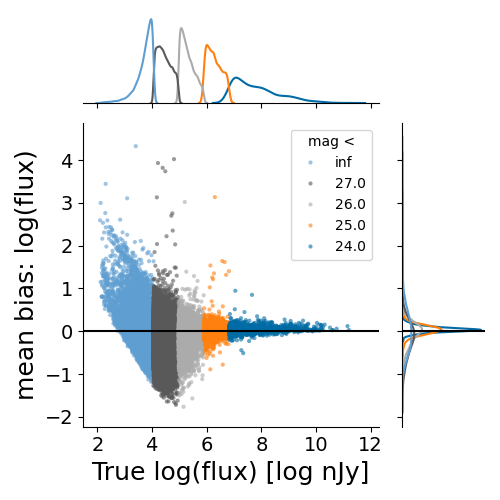}{0.5\textwidth}{(a)}
          \fig{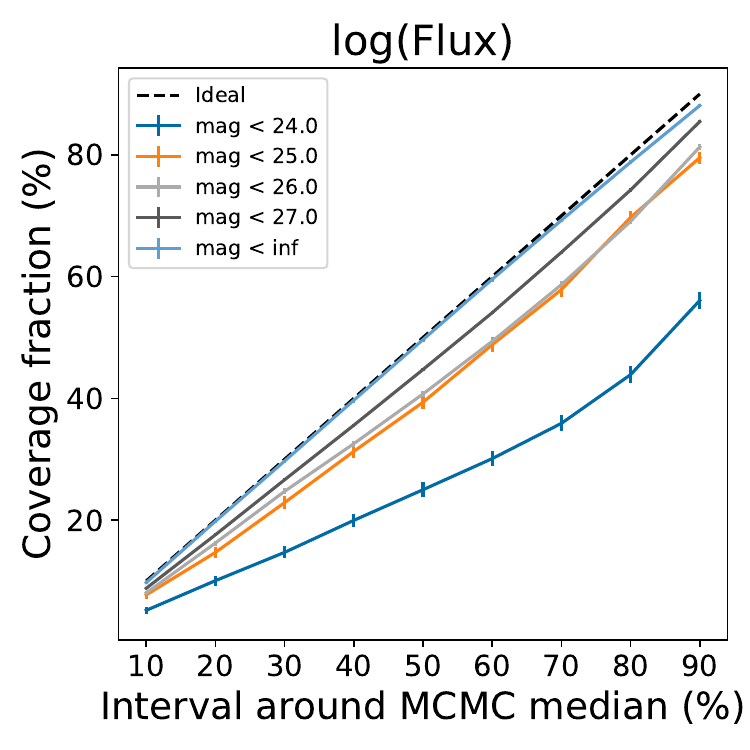}{0.5\textwidth}{(b)}}
\caption{(a): Bias in the log-flux posterior mean. (b): Posterior calibration for log-flux. Realistic data, bulges only.\label{fig:logflux_realistic_bulge}}
\end{figure}

Figure~\ref{fig:intervals_logflux_simplified_bulge}(a) shows a common feature of the interval size plots: as the true galaxy brightness increases, the posterior intervals shift from a steady, flat mag-dependence in the high-mag range, to a region of shrinking and convergence toward zero in the low-mag range. Figures~\ref{fig:hlr_e1_simplified_bulge}(a) and (b) illustrate this tendency in two other parameters of interest: the half-light radius and the first component of ellipticity. These figures show that estimating a parameter from data can require different amounts of galaxy brightness, depending on the parameter. Overall flux takes relatively little galaxy brightness to infer reliably, half-light radius requires a bit more brightness to infer, and ellipticity still more. Our datasets did not contain very many galaxies bright enough to constrain any PSF parameters by an appreciable amount.

For ellipticity, the intervals increase somewhat in the transition region between high-mag and mid-mag, before collapsing rapidly at lower mags. This seems to be caused by the fact that the ellipticity prior is markedly wider in the middle-flux range, as shown by the mountainous shape of the e vs.\ log-flux distribution in Figure~\ref{fig:bulges_2d}. The flat portions of Figures~\ref{fig:hlr_e1_simplified_bulge}(a) and (b) reveal the galaxy brightness regime over which parameter inference is effectively independent of the truth. Figures~\ref{fig:hlr_e1_simplified_bulge}(c) and (d) display corresponding behavior.
\begin{figure}
\gridline{\fig{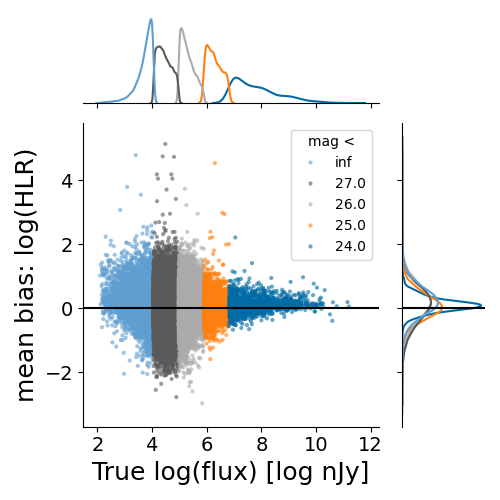}{0.5\textwidth}{(a)}
          \fig{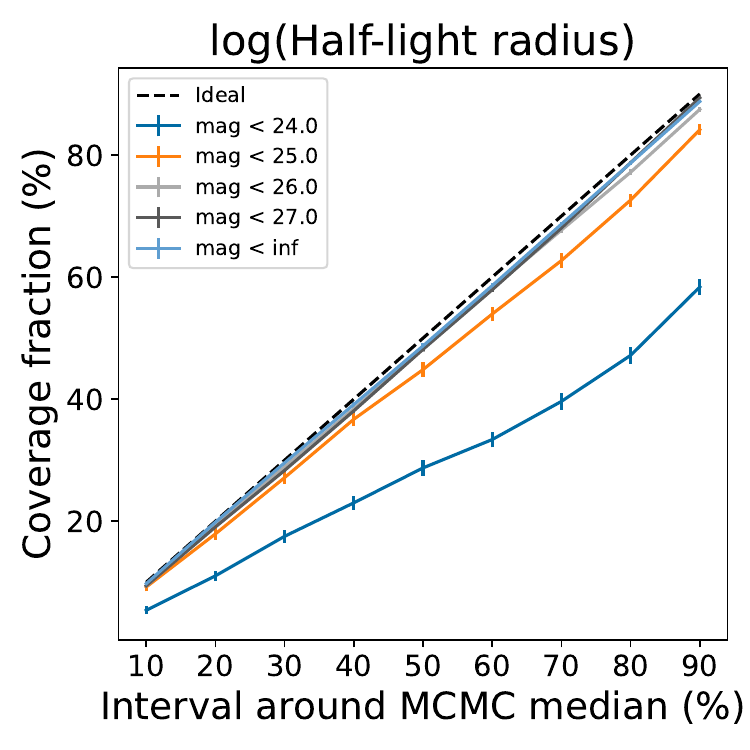}{0.5\textwidth}{(b)}}
\caption{(a): Bias in the log-half-light radius posterior mean. (b): Posterior calibration for log-half-light radius. Realistic data, bulges only.\label{fig:loghlr_realistic_bulge}}
\end{figure}

\subsection{Comparison between Datasets}\label{sec:realistic_results}
Figures~\ref{fig:logflux_simplified_disc}(a) and (b) show log-flux distributions for the simplified disks, as opposed to bulges. The disk plots generally just look like slightly tweaked versions of the bulge plots, with better calibration and somewhat tighter intervals. These differences stem from the fact that disks in our dataset are brighter on average than bulges.
\begin{figure}
\gridline{\fig{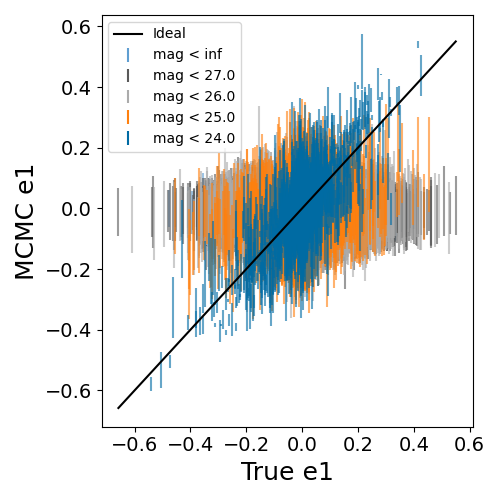}{0.5\textwidth}{(a)}
          \fig{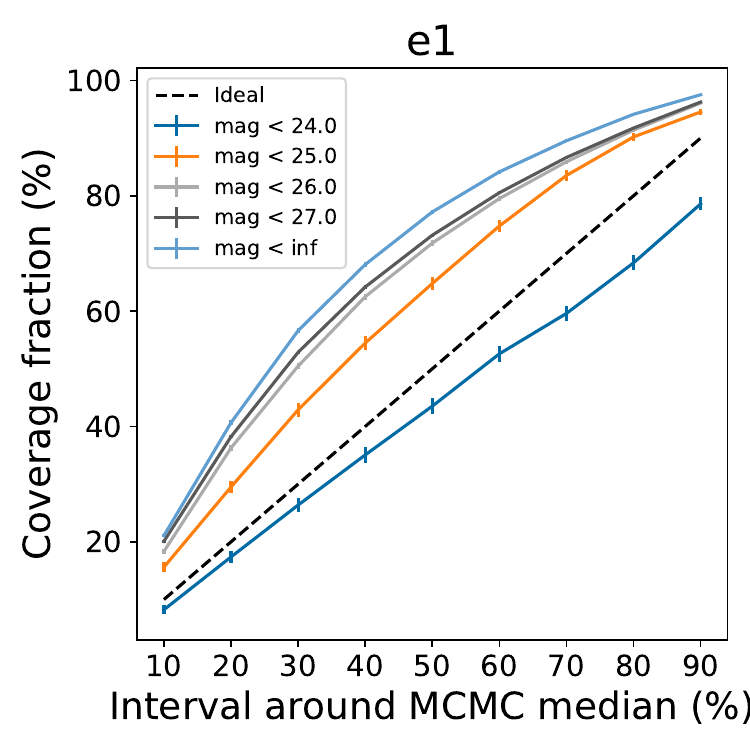}{0.5\textwidth}{(b)}}
\caption{(a): Inner 68\% posterior intervals for e1, plotted vs.\ truth. (b): Posterior calibration for e1. Realistic data, bulges only.\label{fig:e1_realistic_bulge}}
\end{figure}

The realistic data results are mostly rather similar to those of the simplified data, with some important differences. Figure~\ref{fig:logflux_realistic_bulge}(a) shows one of the most noteworthy: a pronounced upward bias in the estimated flux for the brightest galaxies. This bias results in the undercoverage of the flux posteriors for bright galaxies illustrated in Figure~\ref{fig:logflux_realistic_bulge}(b). We see similar trends for the half-light radius as shown in Figure~\ref{fig:loghlr_realistic_bulge}, though generally with better coverage until the brightest magnitudes. The origin of this bias is unknown. Since the likelihood most strongly constrains the posterior for the brightest galaxies, we focus on things that affect the likelihood. Because the issue is with the brightest galaxies, we can ignore the detection corrections to the likelihood.

Figure~\ref{fig:e1_realistic_bulge} shows a bias in ellipticity for the brightest galaxies, again causing undercoverage. Below 24th magnitude, as galaxies get brighter, the coverage steadily gets worse. By the very brightest examples, the measured ellipticity is many sigma away from the truth, as illustrated by the tiny blue lines in Figure~\ref{fig:e1_realistic_bulge}(a) that lie far from the black diagonal. We can say with confidence that we are missing something important in our likelihood model for the realistic data, but we cannot say with confidence what the most important thing is.

We are only fitting galaxies that are strongly bulge-like or disk-like, using matching models in each case, so if the general morphological model is to blame, it is presumably due to the random knots. Knots are not highly pronounced in most DC2 galaxies, however, and we consider other possibilities. The PSF model could be to blame, as an elliptical Kolmogorov form does not exactly match the effective coadd PSF. As noted in Appendix~\ref{sec:psf}, we made an effort to make the PSF prior as correct as possible for this parametric form, but the topic merits additional study.

Finally, correlations between neighboring pixels in coadds, which we do not model here but that we know to be at roughly the 10\% level for immediate neighbors, remain as a stand-out candidate for further study. If this turns out to be the main culprit, it would clearly motivate improvements to the noise modeling in SSI on LSST coadds, because the Pipelines \texttt{source\_injection} module currently assumes uncorrelated pixels. Along these lines, a comparison of SSI on coadds and on separate exposures might help identify the root cause of what we are seeing.

The other most noteworthy difference is the downward shift in the posteriors for the lowest-flux galaxies, compared to the simplified dataset (figure~\ref{fig:bias_logflux_simplified_bulge}a). We attribute this to the selection corrections (section~\ref{sec:conditioning_on_selection}), which give added weight in the likelihood to flux levels below the detection threshold. The size of this downward shift illustrates the significant impact of applying these selection corrections when modeling low-S/N galaxies. The convergence of the low-flux posterior means toward the truth, together with the improvement in calibration shown by the light-blue curve in Figure~\ref{fig:logflux_realistic_bulge}(b), demonstrate the utility of applying these corrections, and they furthermore validate the approximations described in Section~\ref{sec:conditioning_on_selection}.

\section{Conclusions and Connections}\label{sec:conclusions}
\subsection{Summary of Results}
We have developed a framework for MCMC inference of posterior distributions of galaxy properties in LSST images, and tested it on cleanly isolated galaxies detected in $i$-band coadds from the DC2 simulations. Our analysis of 116,088 detection footprints from realistic images establishes that the current \texttt{JIF} framework can be used for large parallel computing runs and can handle most examples of data seen in practice without catastrophic errors.

\citet{2023MNRAS.AstroPhot} contended that ``it is common for [MCMC] uncertainties to underestimate the true uncertainty, especially for fits to real data where the fitted models (i.e., Sérsic) are only approximations to the true shape of a galaxy.'' That claim is motivated with the demonstration of a single simulated example with a posterior that fails to cover the truth within its 95\% credible intervals. Here, we have tested this claim more extensively by looking at the average posterior calibration across more than $10^5$ galaxies.

Our plots in Section~\ref{sec:results} show that---with properly informative priors, a sufficiently well-constructed likelihood including relevant selection corrections, sufficiently long and well-converged MCMC chains, and an assumed Sérsic model at least broadly similar to the true light profile---we can in many cases establish good probabilistic calibration. This means that the posterior intervals can be treated as reliable estimates of uncertainty, across galaxy types and parameter combinations, for at least those galaxies that do not exhibit systematic biases in their posteriors as noted above.

The flux biases we observe in the ``realistic'' data have a simple enough form that one could readily apply a post hoc correction to eliminate the bias and improve coverage. A single overall multiplicative calibration factor would be roughly adequate. The ellipticity biases, affecting the two components of ellipticity, would require at least one additional calibration component to correct. As noted in Section~\ref{sec:intro}, one of the appealing features of Bayesian forward modeling is the prospect of not needing to apply a separate calibration step along with its attendant systematic uncertainties. Our analysis of the ``simplified'' dataset indicates that this promise can be fulfilled if we understand the data-generating process sufficiently well. Biases in the measurements on realistic data demonstrate that we are still missing something significant in our model of the real data-generating process.

One avenue for improvement would be to account for inter-pixel covariance. We could in principle avoid modeling coaddition-induced pixel covariance directly by instead fitting all of the individual exposures of an astrophysical object, like in the application of \texttt{LensFit} to Canada--France--Hawaii Telescope data~\citep{2013MNRAS.429.4}. The latter approach would, in principle, yield the most constraining posteriors, since it doesn't ``throw away data,'' but it multiplies the computational burden, since each separate exposure must be simulated when computing the likelihood. Alternatively, we could accept that even in a Bayesian framework, imperfectly modeled aspects of data generation may still demand post hoc calibration at the end of the day.

The Bayesian approach hinges crucially on the choice of prior, particularly for low-flux footprints, for which the likelihood has little constraining power. This issue is especially acute, as the dimmest footprints will be the most numerous in our data. Furthermore, these are exactly the galaxies that are generally to be detected for the first time by LSST, making it difficult to rigorously establish a prior model for their distribution. Nevertheless, precursor telescope surveys, though they record a substantially smaller fraction of the sky at equivalent depth compared to LSST, still give us some direct empirical sense of the distribution of dim galaxies, which we can use to shape our prior. This was the approach used to the construct the CosmoDC2 simulation, which was verified against the surveys noted in Section~\ref{sec:imagesim}; in the future, deeper and more extensive data from other surveys can be used.

Such data can form the fitting target of a generative model that serves as a prior. Here we have used a BGMM as such a model; other popular approaches such as variational autoencoders \citep{2019FTML.12.4} or normalizing flows \citep{2021IEEE.43.11} could be examined. As we discuss in Section~\ref{sec:mcmc}, our MCMC chains nearly always run to completion using the prior adopted in this study, whereas a more uniform prior will routinely result in fit failures of the kind reported in \citet{2017MNRAS.466.2}, especially on low-S/N galaxy images.

A more realistic prior would have not just a more realistic distribution of galaxy parameters but a more realistic parameterization. By testing our pure bulge and disk models on ``very bulge-like'' and ``very disk-like'' examples that nevertheless contain some fraction of a second (bulge or disk) component, along with knots, our analysis of the realistic dataset includes some model misspecification effects. Even so, we expect that this gives a rather optimistic baseline, which could be made somewhat more realistic by, e.g., adopting a two-component bulge--disk model with variable bulge fraction and fitting this to all detected galaxies. Following \citet{2010MNRAS.404..458V}, each of the components should ideally be allowed a separate ellipticity. We avoided this parameterization in the present study due to concerns about the sharply bimodal prior that would result, which can present problems for some MCMC samplers, but various MCMC methods exist that can handle such priors well~\citep{2025arXiv250105908L}.

\citet{2014MNRAS.441.3} found that for shear measurements, noise bias has a significant interaction with model bias. Although they go on to assert that ``[Bayesian] methods are considered to be free of noise and model related biases,'' we demonstrate here that it is nontrivial to treat either pixel-level variance or galaxy morphology in an entirely satisfactory manner even in a Bayesian framework. Meanwhile, \citet{2017AandA.604.A109} demonstrated that when the prior does not match the true underlying distribution, a noise-dependent ellipticity bias can result even in a fully Bayesian analysis, which they call the ``prior bias.'' It remains to be seen how much this actually impacts a fully Bayesian estimate of shear via hierarchical inference.

\subsection{Connections to Hierarchical Inference}
Before a galaxy can be measured, it must be detected, roughly located in space, deblended, and classified as a galaxy rather than, say, a star. The present study relies on the Pipelines to produce a ``best guess'' for each of these and then applies purity cuts to select just those examples for which the guesses were right. A full analysis must take into account other plausible explanations for what could have produced the image, ideally in a Bayesian fashion.

This is provided by, for example, the BLISS framework, which, given a minimally processed image, produces probabilistic detections of object centroids along with probabilistic star vs.\ galaxy classification, among other outputs~\citep{Mendoza2026.BLISS}. After selecting a different galaxy selection procedure, our selection corrections to the image likelihood will need to be rederived. The correction at this stage does not directly account for spatially correlated shear-dependent selection effects, which will need to be incorporated into the importance sampling weights in subsequent stages of hierarchical inference. Note that in a fully Bayesian hierarchical shear inference scheme, \textit{all} detected galaxies may be used at all stages; if the low-S/N galaxies are modeled well (which we have endeavored to do here), there is no statistical benefit to focusing on a ``gold'' sample specifically for shear analysis.

Usefully, BLISS is an effective deblender on realistically dense images. We have left the important topic of blending to future study. Very roughly, doubling the number of galaxies in a footprint model would double the amount of time it takes to render each light profile (in the likelihood computation) and to compute the prior (once separately for each galaxy). An $n$-galaxy fit would require $9n$ parameters per band to be jointly modeled under our current approach. The affine-invariant ensemble sampling strategies described in G\&W do not require any separate tuning for separate parameters and should scale gracefully to such fits. For footprints without too much blending-induced ambiguity, the computational burden should be roughly equivalent to modeling each galaxy separately.

Our method appears to be competitive in speed compared to several other published approaches, as noted in Section~\ref{subsec:chain_properties}. Nevertheless, if exactly the present method is scaled up from 2.4 square degrees to 18,000 square degrees---the expected total sky coverage of LSST \citep{2019ApJ.873.2}---60 million CPU hours would be required. Modeling all six filter bands would multiply this burden by an additional factor (probably less than six, since various steps of the computation remain unchanged even as more bands/images are added), as would any of the other more detailed modeling proposals outlined above, such as simultaneous modeling of blended galaxies. As shown below, however, the actual time requirement for effective hierarchical inference may be considerably less onerous than this.

The present study treats each galaxy as fully independent from all others and makes no attempt at the kind of correlated analysis necessary for weak lensing inference. Under the framework of \citet{2015ApJ.807.1}, this can happen in a subsequent step of hierarchical inference, following probabilistic sampling of galaxy catalogs and the properties of each galaxy. We stress that the posterior samples obtained on individual galaxy parameters would serve here as provisional, ``interim'' posteriors. From this standpoint, the most essential requirements on single-galaxy fits are that the MCMC chains converge and run for long enough to thoroughly sample the interim posterior. \citet{Mendoza2026.shear} demonstrated exactly how this may be done, on significantly simplified simulated LSST-like images.

They showed that 300 effective MCMC samples of galaxy properties are sufficient to achieve a multiplicative shear bias $|m|$ below the LSST requirement of $2 \times 10^{-3}$, and with differentiable code implementations, gradient-based samplers, and GPUs, the sampling can be done within 0.45 seconds per galaxy. Furthermore, they demonstrated that an exact knowledge of intrinsic galaxy properties (the basis of our ``informative prior''), while useful, is not strictly necessary to achieve a low $|m|$ so long as the image likelihood is very well characterized and that the PSF and sky background are known.

Our study complements this work by sharpening our understanding of the LSST coadded image likelihood, as well as by including PSF size and shape as additional nuisance parameters in the probabilistic model. With a synthesis of the preceding methods, we therefore think it is feasible to implement a fully Bayesian end-to-end cosmic shear inference pipeline for LSST, with MCMC-sampled galaxy measurements as a component. Using the highly efficient Gaussian process method of \citet{Sallaberry2025.981.2}, this can be extended a step further to posterior inference of the lensing potential across wide swaths of the sky.

\begin{acknowledgments}
Scott Perkins (LLNL) provided insight into the role of detection in conditioning the likelihood. This work was performed under the auspices of the U.S.
Department of Energy (DOE) by Lawrence Livermore National Laboratory (LLNL) under Contract DE-AC52-07NA27344, with IM release No.\ LLNL-JRNL-853846.
Funding for this work was provided as part of the DOE Office of Science, High Energy Physics cosmic frontier program.

This document was prepared as an account of work sponsored by an agency of
the United States government. Neither the United States government nor Lawrence
Livermore National Security, LLC, nor any of their employees makes any warranty,
expressed or implied, or assumes any legal liability or responsibility for the accuracy,
completeness, or usefulness of any information, apparatus, product, or process disclosed,
or represents that its use would not infringe privately owned rights. Reference
herein to any specific commercial product, process, or service by trade name, trademark,
manufacturer, or otherwise does not necessarily constitute or imply its endorsement,
recommendation, or favoring by the United States government or Lawrence Livermore
National Security, LLC. The views and opinions of authors expressed herein
do not necessarily state or reflect those of the United States government or Lawrence
Livermore National Security, LLC, and shall not be used for advertising or product
endorsement purposes.
\end{acknowledgments}

\software{
        \texttt{JIF} (\href{https://github.com/mdschneider/JIF}{\nolinkurl{github.com/mdschneider/JIF}}),
        \texttt{GCRCatalogs} \citep{2018APJS.234.2},
        \texttt{GalSim} (\href{https://galsim-developers.github.io/GalSim/_build/html/index.html}{\nolinkurl{galsim-developers.github.io/GalSim}}),
        LSST Science Pipelines (\href{https://pipelines.lsst.io}{\nolinkurl{pipelines.lsst.io}}),
        \texttt{scikit-learn} (\href{https://scikit-learn.org/stable/}{\nolinkurl{scikit-learn.org}}),
        \texttt{emcee} (\href{https://emcee.readthedocs.io/en/stable/}{\nolinkurl{emcee.readthedocs.io}}),
        \texttt{scipy} (\href{https://docs.scipy.org/doc/scipy/}{\nolinkurl{docs.scipy.org}})
}

\bibliography{draft}{}

@misc{rtn-011,
	author = {Leanne Guy and Yusra {AlSayyad} and Keith Bechtol
	and Eric Bellm and Robert Blum and Gregory Dubois-Felsmann
	and Frossie Economou and Melissa Graham and Željko Ivezić
	and R. Lynne Jones and Robert Lupton and Phil Marshall
	and William {O’Mullane} and Colin Slater and Michael Strauss},
	title = {Rubin Plans for an Early Science Program},
	month = {jul},
	year = {2026},
	url = {https://rtn-011.lsst.io}
}

@misc{jif,
	author = {Michael Schneider and
	James Buchanan and
	Will Dawson and
	Kerianne Pruett},
	title = {JIF},
	publisher = {Zenodo},
	month = {aug},
	year = {2025},
	doi = {10.5281/zenodo.16781514},
	url = {https://doi.org/10.5281/zenodo.16781514}
}

@incollection{2025arXiv250105908L,
  author    = {Łatuszyński, Krzysztof and Moores, Matthew T. and Stumpf-Fétizon, Timothée},
  title     = {MCMC Methods for Multi-Modal Distributions},
  booktitle = {Handbook of Markov Chain Monte Carlo},
  editor    = {Craiu, Radu V. and Vats, Deepak and Jones, G. L. and Brooks, Steve and Gelman, Andrew and Meng, Xiao-Li},
  edition   = {2},
  year      = {2026},
  publisher = {Chapman and Hall/CRC},
  doi       = {10.1201/9781003453420},
  url       = {https://doi.org/10.1201/9781003453420}
}

@article{2010MNRAS.404..458V,
       author = {{Voigt}, L.~M. and {Bridle}, S.~L.},
        title = "{Limitations of model-fitting methods for lensing shear estimation}",
      journal = {\mnras},
         year = 2010,
        month = may,
       volume = {404},
       number = {1},
        pages = {458-467},
          doi = {10.1111/j.1365-2966.2010.16300.x},
archivePrefix = {arXiv},
       eprint = {0905.4801},
 primaryClass = {astro-ph.CO},
       adsurl = {https://ui.adsabs.harvard.edu/abs/2010MNRAS.404..458V}
}

@article{2025MNRAS.543.4156Y,
       author = {{Yamamoto}, M. and {Becker}, M.~R. and {Sheldon}, E. and {Jarvis}, M. and {Gruendl}, R.~A. and {Menanteau}, F. and {Rykoff}, E.~S. and {Mau}, S. and {Schutt}, T. and {Gatti}, M. and {Troxel}, M.~A. and {Amon}, A. and {Anbajagane}, D. and {Bernstein}, G.~M. and {Gruen}, D. and {Huff}, E.~M. and {Tabbutt}, M. and {Tong}, A. and {Yanny}, B. and {Abbott}, T.~M.~C. and {Aguena}, M. and {Alarcon}, A. and {Andrade-Oliveira}, F. and {Bechtol}, K. and {Blazek}, J. and {Brooks}, D. and {Rosell}, A. Carnero and {Carretero}, J. and {Chang}, C. and {Choi}, A. and {Costanzi}, M. and {Crocce}, M. and {da Costa}, L.~N. and {Davis}, T.~M. and {De Vicente}, J. and {Desai}, S. and {Diehl}, H.~T. and {Dodelson}, S. and {Doel}, P. and {Doux}, C. and {Drlica-Wagner}, A. and {Fert{\'e}}, A. and {Flaugher}, B. and {Fosalba}, P. and {Frieman}, J. and {Garc{\'\i}a-Bellido}, J. and {Gaztanaga}, E. and {Giannini}, G. and {Gutierrez}, G. and {Hartley}, W.~G. and {Herner}, K. and {Hinton}, S.~R. and {Hollowood}, D.~L. and {Honscheid}, K. and {Huterer}, D. and {Krause}, E. and {Kuehn}, K. and {Lahav}, O. and {Lima}, M. and {Marshall}, J.~L. and {Mena-Fern{\'a}ndez}, J. and {Miquel}, R. and {Mohr}, J.~J. and {Muir}, J. and {Myles}, J. and {Ogando}, R.~L.~C. and {Pieres}, A. and {Malag{\'o}n}, A.~A. Plazas and {Porredon}, A. and {Prat}, J. and {Raveri}, M. and {Rodriguez-Monroy}, M. and {Roodman}, A. and {Samuroff}, S. and {Sanchez}, E. and {Sanchez Cid}, D. and {Scarpine}, V. and {Sevilla-Noarbe}, I. and {Smith}, M. and {Soares-Santos}, M. and {Suchyta}, E. and {Tarle}, G. and {Vikram}, V. and {Weaverdyck}, N. and {Wiseman}, P. and {Zhang}, Y.},
        title = "{Dark Energy Survey Year 6 results: cell-based coadds and METADETECTION weak lensing shape catalogue}",
      journal = {\mnras},
         year = 2025,
        month = nov,
       volume = {543},
       number = {4},
        pages = {4156-4186},
          doi = {10.1093/mnras/staf1661},
archivePrefix = {arXiv},
       eprint = {2501.05665},
 primaryClass = {astro-ph.CO},
       adsurl = {https://ui.adsabs.harvard.edu/abs/2025MNRAS.543.4156Y}
}

@article{2023OJAp.6,
	author = {Sheldon, Erin and Becker, Matthew and Jarvis, Michael and Armstrong, Robert and {the LSST Dark Energy Science Collaboration}},
	doi = {10.21105/astro.2303.03947},
	journal = {OJAp},
	month = {May},
	year = {2023},
	title = {{Metadetection Weak Lensing for the Vera C. Rubin Observatory}},
	volume = {6},
}

@article{Park2025.537.1,
	author = {Park, Andy and Li, Xiangchong and Mandelbaum, Rachel},
	doi = {10.1093/mnras/staf053},
	journal = {\mnras},
	month = {feb},
	year = {2025},
	title = {{Accurate shear estimation with fourth-order moments}},
	volume = {537},
	number = {1},
    pages = {507},
}

@article{Li2024.527.4,
	author = {Li, Xiangchong and Mandelbaum, Rachel and Jarvis, Mike and Li, Yin and Park, Andy and Zhang, Tianqing},
	doi = {10.1093/mnras/stad3895},
	journal = {\mnras},
	month = {feb},
	year = {2024},
	title = {{A differentiable perturbation-based weak lensing shear estimator}},
	volume = {527},
	number = {4},
	pages = {10388}
}

@article{Li2022.511.4,
	author = {Li, Xiangchong and Li, Yin and Massey, Richard},
	doi = {10.1093/mnras/stac342},
	journal = {\mnras},
	month = {apr},
	year = {2022},
	title = {{Weak gravitational lensing shear measurement with FPFS: analytical mitigation of noise bias and selection bias}},
	volume = {511},
	number = {4},
	pages = {4850}
}

@article{Li2018.481.4,
	author = {Li, Xiangchong and Katayama, Nobuhiko and Oguri, Masamune and More, Surhud},
	doi = {10.1093/mnras/sty2548},
	journal = {\mnras},
	month = {dec},
	year = {2018},
	title = {{Fourier Power Function Shapelets (FPFS) shear estimator: performance on image simulations}},
	volume = {481},
	number = {4},
	pages = {4445}
}

@article{Li2025.jun,
	author = {Li, Xiangchong},
	doi = {10.48550/arXiv.2506.16607},
	eprint = {2506.16607},
	month = {jun},
	year = {2025},
	title = {{Analytical weak-lensing shear response of galaxy model fitting}},
}

@misc{eve_kovacs_2022_6336066,
  author       = {Eve Kovacs and
                  Rachel Mandelbaum and
                  Mike Jarvis and
                  Chris Walter},
  title        = {Ellipticity Distribution Bug in cosmoDC2 and
                   SkySim5000\_v1.1.1
                  },
  month        = mar,
  year         = 2022,
  publisher    = {Zenodo},
  doi          = {10.5281/zenodo.6336066},
  url          = {https://doi.org/10.5281/zenodo.6336066},
}

@article{guerrini2024,
    author = {{Guerrini, Sacha} and {Kilbinger, Martin} and {Leterme, Hubert} and {Guinot, Axel} and {Wang, Jingwei} and {Peters, Fabian Hervas} and {Hildebrandt, Hendrik} and {Hudson, Michael J.} and {McConnachie, Alan}},
    title = {Galaxy–point spread function correlations as a probe of weak-lensing systematics with UNIONS data},
    DOI= "10.1051/0004-6361/202453512",
    url= "https://doi.org/10.1051/0004-6361/202453512",
    journal = {A\&A},
    year = 2025,
    volume = 700,
    pages = "A215",
    eprint={2412.14666},
    archivePrefix={arXiv},
    primaryClass={astro-ph.CO},
}

@article{Lamman2024IA,
	author = {Lamman, Claire and Tsaprazi, Eleni and Shi, Jingjing and {\v S}ar{\v c}evi{\' c}, Nikolina Niko and Pyne, Susan and Legnani, Elisa and Ferreira, Tassia},
	journal = {OJAp},
	doi = {10.21105/astro.2309.08605},
	year = {2024},
	month = {feb 23},
	publisher = {Maynooth Academic Publishing},
	title = {The {IA} {Guide}: A {Breakdown} of {Intrinsic} {Alignment} {Formalisms}},
	volume = {7},
}

@article{li2023hypersuprimecamyear3,
    title = {Hyper Suprime-Cam Year 3 results: Cosmology from cosmic shear two-point correlation functions},
    author = {Li, Xiangchong and Zhang, Tianqing and Sugiyama, Sunao and Dalal, Roohi and Terasawa, Ryo and Rau, Markus M. and Mandelbaum, Rachel and Takada, Masahiro and More, Surhud and Strauss, Michael A. and Miyatake, Hironao and Shirasaki, Masato and Hamana, Takashi and Oguri, Masamune and Luo, Wentao and Nishizawa, Atsushi J. and Takahashi, Ryuichi and Nicola, Andrina and Osato, Ken and Kannawadi, Arun and Sunayama, Tomomi and Armstrong, Robert and Bosch, James and Komiyama, Yutaka and Lupton, Robert H. and Lust, Nate B. and MacArthur, Lauren A. and Miyazaki, Satoshi and Murayama, Hitoshi and Nishimichi, Takahiro and Okura, Yuki and Price, Paul A. and Tait, Philip J. and Tanaka, Masayuki and Wang, Shiang-Yu},
    journal = {PhRvD},
    volume = {108},
    issue = {12},
    pages = {123518},
    numpages = {39},
    year = {2023},
    month = {Dec},
    publisher = {American Physical Society},
    doi = {10.1103/PhysRevD.108.123518},
    url = {https://link.aps.org/doi/10.1103/PhysRevD.108.123518}
}

@article{Anbajagane2025Dark,
	author = {Anbajagane, Dhayaa and Tabbutt, M. and Beas-Gonzalez, J. and Yanny, B. and Everett, S. and Becker, M. R. and Yamamoto, M. and Legnani, E. and De Vicente, J. and Bechtol, K. and Elvin-Poole, J. and Bernstein, G. M. and Choi, A. and Gatti, M. and Giannini, G. and Gruendl, R. A. and Jarvis, M. and Lee, S. and Mena-Fern{\' a}ndez, J. and Porredon, A. and Rodriguez-Monroy, M. and Rozo, E. and Rykoff, E. S. and Schutt, T. and Sheldon, E. and Troxel, M. A. and Weaverdyck, N. and Wetzell, V. and Aguena, M. and Alarcon, A. and Allam, S. and Amon, A. and Andrade-Oliveira, F. and Brooks, D. and Rosell, A. Carnero and Carretero, J. and Chang, C. and Crocce, M. and da Costa, L. N. and Pereira, M. E. S. and Davis, T. M. and Desai, S. and Diehl, H. T. and Dodelson, S. and Doel, P. and Drlica-Wagner, A. and Fert{\' e}, A. and Frieman, J. and Garc{\' i}a-Bellido, J. and Gaztanaga, E. and Gruen, D. and Gutierrez, G. and Hartley, W. G. and Herner, K. and Hinton, S. R. and Hollowood, D. L. and Honscheid, K. and Huterer, D. and James, D. J. and Krause, E. and Kuehn, K. and Lahav, O. and Marshall, J. L. and Miquel, R. and Muir, J. and Myles, J. and Pieres, A. and Malag{\' o}n, A. A. Plazas and Prat, J. and Raveri, M. and Samuroff, S. and Sanchez, E. and Cid, D. Sanchez and Sevilla-Noarbe, I. and Smith, M. and Suchyta, E. and Tarle, G. and Tucker, D. L. and Walker, A. R. and Wiseman, P. and Zhang, Y. and {The DES Collaboration}},
	journal = {OJAp},
	doi = {10.33232/001c.138627},
	year = {2025},
	month = {may 29},
	publisher = {Maynooth Academic Publishing},
	title = {Dark {Energy} {Survey} {Year} 6 {Results}: Synthetic-source {Injection} {Across} the {Full} {Survey} {Using} {Balrog}},
	volume = {8},
}

@article{1996AAPS.117,
	author = {{Bertin}, E. and {Arnouts}, S.},
	doi = {10.1051/aas:1996164},
	journal = {\aaps},
	month = {Jun},
	title = {{SExtractor: Software for source extraction}},
	volume = {117},
    pages = {393},
	year = {1996}
}

@article{2006MNRAS.368.3,
	author = {{Heymans}, Catherine and {Van Waerbeke}, Ludovic and {Bacon}, David and {Berge}, Joel and {Bernstein}, Gary and {Bertin}, Emmanuel and {Bridle}, Sarah and {Brown}, Michael and {Clowe}, Douglas and {Dahle}, Håkon and {Erben}, Thomas and {Gray}, Meghan and {Hetterscheidt}, Marco and {Hoekstra}, Henk and {Hudelot}, Patrick and {Jarvis}, Mike and {Kuijken}, Konrad and {Margoniner}, Vera and {Massey}, Richard and {Mellier}, Yannick and {Nakajima}, Reiko and {Refregier}, Alexandre and {Rhodes}, Jason and {Schrabback}, Tim and {Wittman}, David},
	doi = {10.1111/j.1365-2966.2006.10198.x},
	journal = {\mnras},
	month = {Apr},
	title = {{The Shear Testing Programme – I. Weak lensing analysis of simulated ground-based observations}},
	volume = {368},
	number = {3},
    pages = {1323},
	year = {2006}
}

@misc{2013ascl.soft07006B,
       author = {{Bridle}, S.~L. and {Kneib}, J. -P. and {Bardeau}, S.},
        title = "{im2shape: Bayesian Galaxy Shape Estimation}",
 howpublished = {Astrophysics Source Code Library, record ascl:1307.006},
         year = {2013},
        month = {Jul},
          eid = {ascl:1307.006},
        pages = {ascl:1307.006},
archivePrefix = {ascl},
       eprint = {1307.006},
       adsurl = {https://ui.adsabs.harvard.edu/abs/2013ascl.soft07006B}
}

@article{1995ApJ.449,
	author = {{Kaiser}, Nick and {Squires}, Gordon and {Broadhurst}, Tom},
	doi = {10.1086/176071},
	eprint = {astro-ph/9411005},
	journal = {\apj},
	month = {Aug},
	title = {{A Method for Weak Lensing Observations}},
	volume = {449},
	year = {1995},
    pages = {460},
}

@article{1999AandA.352,
	author = {{Kuijken}, Konrad},
	doi = {10.48550/arXiv.astro-ph/9904418},
	eprint = {astro-ph/9904418},
	journal = {\aap},
	month = {Dec},
	title = {{Weak weak lensing: correcting weak shear measurements accurately for PSF anisotropy}},
	volume = {352},
	year = {1999},
    pages = {355},
}

@inproceedings{2001Bridle,
	author = {{Bridle}, S. and {Kneib}, J.-P. and {Bardeau}, S. and {Gull}, S.},
	doi = {10.1142/9789812778017_0006},
	booktitle = {{Proceedings of the Yale Cosmology Workshop, `The Shapes of Galaxies and Their Dark Matter Halos'}},
	publisher = {World Scientific Press},
	city = {Singapore},
	month = {May},
	title = {{Bayesian Galaxy Shape Estimation}},
	year = {2001}
}

@article{2008MNRAS.390.1,
	author = {{Kitching}, T. and {Miller}, L. and {Heymans}, C. and {van Waerbeke}, L. and {Heavens}, A.},
	doi = {10.1111/j.1365-2966.2008.13628.x},
	eprint = {0802.1528},
	journal = {\mnras},
	month = {Jun},
	title = {{Bayesian galaxy shape measurement for weak lensing surveys - II. Application to simulations}},
	volume = {390},
	number = {1},
	year = {2008},
    pages = {149},
}

@article{2007MNRAS.382.1,
	author = {{Miller}, L. and {Kitching}, T. and {Heymans}, C. and {Heavens}, A. and {Van Waerbeke}, L.},
	doi = {10.1111/j.1365-2966.2007.12363.x},
	eprint = {0708.2340},
	journal = {\mnras},
	month = {Nov},
	title = {{Bayesian galaxy shape measurement for weak lensing surveys – I. Methodology and a fast-fitting algorithm}},
	volume = {382},
	number = {1},
    pages = {315},
	year = {2007}
}

@article{Liang2026.9,
    author = {Liang, Shuang and Adari, Prakruth and von der Linden, Anja and {LSST Dark Energy Science Collaboration}},
    journal = {OJAp},
    doi = {10.33232/001c.162504},
    year = {2026},
    month = {may 25},
    publisher = {Maynooth Academic Publishing},
    title = {Catalog-based detection of unrecognized blends in deep optical ground based imaging},
    volume = {9},
}

@article{Mendoza2026.BLISS,
    author = {Mendoza, Ismael and Hansen, Derek and Liu, Runjing and Zhao, Zhe and Pang, Ziteng and Guinot, Axel and Avestruz, Camille and Regier, Jeffrey and {The LSST Dark Energy Science Collaboration}},
    journal = {OJAp},
    doi = {10.33232/001c.158908},
    year = {2026},
    month = {mar 12},
    publisher = {Maynooth Academic Publishing},
    title = {Simulation-{Based} {Inference} for {Probabilistic} {Galaxy} {Detection} and {Deblending}},
    volume = {9},
}

@article{Mendoza2026.shear,
    author = {Mendoza, Ismael and Guinot, Axel and Becker, Matthew R. and Avestruz, Camille and Campagne, Jean-Eric and Porqueres, Natalia and Schneider, Michael and Tsaprazi, Eleni and {LSST Dark Energy Science Collaboration}},
    journal = {OJAp},
    doi = {10.33232/001c.168201},
    year = {2026},
    month = {aug 24},
    publisher = {Maynooth Academic Publishing},
    title = {Differentiable {Forward} {Modeling} for {Efficient} and {Accurate} {Shear} {Inference}},
    volume = {9},
}

@ARTICLE{Sallaberry2025.981.2,
       author = {{Sallaberry}, Gregory and {Priest}, Benjamin W. and {Armstrong}, Robert and {Schneider}, Michael and {Muyskens}, Amanda and {Steil}, Trevor and {Iwabuchi}, Keita},
        title = "{A Scalable Gaussian Process Approach to Shear Mapping with MuyGPs}",
      journal = {\apj},
         year = 2025,
        month = mar,
       volume = {981},
       number = {2},
          eid = {166},
        pages = {166},
          doi = {10.3847/1538-4357/adb0b7},
archivePrefix = {arXiv},
       eprint = {2410.00308},
 primaryClass = {astro-ph.CO},
       adsurl = {https://ui.adsabs.harvard.edu/abs/2025ApJ...981..166S}
}

@article{2017AandA.604.A109,
    author = {{Simon, Patrick} and {Schneider, Peter}},
    title = {Weak-lensing shear estimates with general adaptive moments,  and studies of bias by pixellation, PSF distortions, and noise},
    DOI= "10.1051/0004-6361/201629591",
    url= "https://doi.org/10.1051/0004-6361/201629591",
    journal = {A\&A},
    year = 2017,
    volume = 604,
    pages = "A109",
}

@article{2017MNRAS.471.1,
	author = {{Tessore}, Nicolas},
	doi = {10.1093/mnrasl/slx100},
	eprint = {1705.01109},
	journal = {\mnras},
	month = {Jul},
	title = {{An unbiased estimator for the ellipticity from image moments}},
	volume = {471},
	number = {1},
    pages = {L57},
	year = {2017}
}

@article{2014MNRAS.441.3,
	author = {{Kacprzak}, Tomasz and {Bridle}, Sarah and {Rowe}, Barnaby and {Voigt}, Lisa and {Zuntz}, Joe and {Hirsch}, Michael and {MacCrann}, Niall},
	doi = {10.1093/mnras/stu588},
	eprint = {1308.4663},
	journal = {\mnras},
	month = {May},
	title = {{Sérsic galaxy models in weak lensing shape measurement: model bias, noise bias and their interaction}},
	volume = {441},
	number = {3},
	year = {2014},
    pages = {2528},
}

@article{2012MNRAS.425.3,
	author = {{Refregier}, Alexandre and {Kacprzak}, Tomasz and {Amara}, Adam and {Bridle}, Sarah and {Rowe}, Barnaby},
	doi = {10.1111/j.1365-2966.2012.21483.x},
	eprint = {1203.5050},
	journal = {\mnras},
	month = {Sep},
	title = {{Noise bias in weak lensing shape measurements}},
	volume = {425},
	number = {3},
    pages = {1951},
	year = {2012}
}

@article{2021NatRevPhys.3,
	author = {{Melchior}, Peter and {Joseph}, Rémy and {Sanchez}, Javier and {MacCrann}, Niall and {Gruen}, Daniel},
	doi = {10.1038/s42254-021-00353-y},
	journal = {NatRP},
	month = {Oct},
	title = {{The challenge of blending in large sky surveys}},
	volume = {3},
    pages = {712},
	year = {2021}
}

@article{2013PASP.125,
	author = {{Foreman-Mackey}, Daniel and {Hogg}, David and {Lang}, Dustin and {Goodman}, Jonathan},
	doi = {10.1086/670067},
	journal = {\pasp},
	month = {Feb},
	title = {{emcee: The MCMC Hammer}},
	volume = {125},
    pages = {306},
	year = {2013}
}

@article{2010CAMCS.5.1,
	author = {{Goodman}, Jonathan and {Weare}, Jonathan},
	doi = {10.2140/camcos.2010.5.65},
	journal = {Comm.~App.~Math.~and~Comp.~Sci.},
	month = {Jan},
	title = {{Ensemble samplers with affine invariance}},
	volume = {5},
	number = {1},
    pages = {65},
	year = {2010}
}

@article{2010ApJS.191,
	author = {{Spergel}, David},
	doi = {10.1088/0067-0049/191/1/58},
	eprint = {1007.3248},
	journal = {\apjs},
	month = {Oct},
	title = {{Analytical Galaxy Profiles for Photometric and Lensing Analysis}},
	volume = {191},
    pages = {58},
	year = {2010}
}

@article{2001PhysRep.340,
	author = {{Bartelmann}, Matthias and {Schneider}, Peter},
	doi = {10.1016/S0370-1573(00)00082-X},
	eprint = {astro-ph/9912508},
	journal = {\physrep},
	month = {Jan},
	title = {{Weak gravitational lensing}},
	volume = {340},
    pages = {291},
	year = {2001}
}

@article{2019ApJ.873.2,
	archiveprefix = {arXiv},
	author = {{Ivezi{\'c}}, {\v Z}eljko and {Kahn}, Steven and {Tyson}, J. Anthony and others},
	doi = {10.3847/1538-4357/ab042c},
	eprint = {0805.2366},
	journal = {\apj},
	month = {Mar},
	number = {2},
	title = {{LSST: From Science Drivers to Reference Design and Anticipated Data Products}},
	volume = {873},
    pages = {111},
	year = {2019}
}

@article{2018ARAA.56,
	archiveprefix = {arXiv},
	author = {{Mandelbaum}, Rachel},
	doi = {10.1146/annurev-astro-081817-051928},
	eprint = {1710.03235},
	journal = {\araa},
	month = {Sep},
	title = {{Weak Lensing for Precision Cosmology}},
	volume = {56},
    pages = {393},
	year = {2018}
}

@article{2018PASJ.70.SP1.a,
	archiveprefix = {arXiv},
	author = {{Bosch}, James and {Armstrong}, Robert and {Bickerton}, Steven and {Furusawa}, Hisanori and {Ikeda}, Hiroyuki and {Michitaro}, Koike and {Lupton}, Robert and {Mineo}, Sogo and {Price}, Paul and {Takata}, Tadafumi and {Tanaka}, Masayuki and {Yasuda}, Naoki and {AlSayyad}, Yusra and {Becker}, Andrew and {Coulton}, William and {Coupon}, Jean and {Garmilla}, Jose and {Huang}, Song and {Krughoff}, K. Simon and {Lang}, Dustin and {Leauthaud}, Alexie and {Lim}, Kian-Tat and {Lust}, Nate and {MacArthur}, Lauren and {Mandelbaum}, Rachel and {Miyatake}, Hironao and {Miyazaki}, Satoshi and {Murata}, Ryoma and {More}, Surhud and {Okura}, Yuki and {Owen}, Russell and {Swinbank}, John and {Strauss}, Michael and {Yamada}, Yoshihiko and {Yamanoi}, Hitomi},
	doi = {10.1093/pasj/psx080},
	eprint = {1705.06766},
	journal = {\pasj},
	month = {Jan},
	number = {SP1},
	title = {{The Hyper Suprime-Cam Software Pipeline}},
	volume = {70},
    pages = {S5},
	year = {2018}
}

@article{2019APJS.245.2,
	archiveprefix = {arXiv},
	author = {{Korytov}, Danila and {Hearin}, Andrew and {Kovacs}, Eve and {Larsen}, Patricia and {Rangel}, Esteban and {Hollowed}, Joseph and {Benson}, Andrew and {Heitmann}, Katrin and {Mao}, Yao-Yuan and {Bahmanyar}, Anita and {Chang}, Chihway and {Campbell}, Duncan and {DeRose}, Joseph and {Finkel}, Hal and {Frontiere}, Nicholas and {Gawiser}, Eric and {Habib}, Salman and {Joachimi}, Benjamin and {Lanusse}, Fran{\c c}ois and {Li}, Nan and {Mandelbaum}, Rachel and {Morrison}, Christopher and {Newman}, Jeffrey and {Pope}, Adrian and {Rykoff}, Eli and {Simet}, Melanie and {To}, Chun-Hao and {Vikraman}, Vinu and {Wechsler}, Risa and {White}, Martin},
	doi = {10.3847/1538-4365/ab510c},
	eprint = {1907.06530},
	journal = {\apjs},
	month = {Dec},
	number = {2},
	title = {{CosmoDC2: A Synthetic Sky Catalog for Dark Energy Science with LSST}},
	volume = {245},
	year = {2019},
    pages = {26},
}

@article{2021APJS.253.1,
	archiveprefix = {arXiv},
	author = {{Abolfathi}, Bela and {Alonso}, David and {Armstrong}, Robert and others},
	doi = {10.3847/1538-4365/abd62c},
	eprint = {2010.05926},
	journal = {\apjs},
	month = {Mar},
	number = {1},
	title = {{The LSST DESC DC2 Simulated Sky Survey}},
	volume = {253},
    pages = {31},
	year = {2021}
}

@article{2012_DESC,
	archiveprefix = {arXiv},
	author = {{LSST Dark Energy Science Collaboration}},
	eprint = {1211.0310},
	month = {Nov},
	year = {2012}
}

@article{2021_DESCDC2_datarelease,
	archiveprefix = {arXiv},
	author = {{LSST Dark Energy Science Collaboration}},
	eprint = {2101.04855},
	month = {Jan},
	title = {{DESC DC2 Data Release Note}},
	year = {2021}
}

@article{2019APJS.245.1,
	archiveprefix = {arXiv},
	author = {{Heitmann}, Katrin and {Finkel}, Hal and {Pope}, Adrian and {Morozov}, Vitali and {Frontiere}, Nicholas and {Habib}, Salman and {Rangel}, Esteban and {Uram}, Thomas and {Korytov}, Danila and {Child}, Hillary and {Flender}, Samuel and {Insley}, Joe and {Rizzi}, Silvio},
	doi = {10.3847/1538-4365/ab4da1},
	eprint = {1904.11970},
	journal = {\apjs},
	month = {Nov},
	number = {1},
	title = {{The Outer Rim Simulation: A Path to Many-core Supercomputers}},
	volume = {245},
    pages = {16},
	year = {2019}
}

@article{2011APJS.192.2,
	archiveprefix = {arXiv},
	author = {{Komatsu}, E. and {Smith}, K. and {Dunkley}, J. and {Bennett}, C. and {Gold}, B. and {Hinshaw}, G. and {Jarosik}, N. and {Larson}, D. and {Nolta}, M. and {Page}, L. and {Spergel}, D. and {Halpern}, M. and {Hill}, R. and {Kogut}, A. and {Limon}, M. and {Meyer}, S. and {Odegard}, N. and {Tucker}, G. and {Weiland}, J. and {Wollack}, E. and {Wright}, E.},
	doi = {10.1088/0067-0049/192/2/18},
	eprint = {1001.4538},
	journal = {\apjs},
	month = {Jan},
	number = {2},
	title = {{Seven-Year Wilkinson Microwave Anisotropy Probe (WMAP) Observations: Cosmological Interpretation}},
	volume = {192},
	year = {2011},
    pages = {18},
}

@article{2018PASJ.70.SP1.b,
	archiveprefix = {arXiv},
	author = {{Aihara}, Hiroaki and {Armstrong}, Robert and {Bickerton}, Steven and others},
	doi = {10.1093/pasj/psx081},
	eprint = {1702.08449},
	journal = {PASJ},
	month = {Jan},
	number = {SP1},
	title = {{First data release of the Hyper Suprime-Cam Subaru Strategic Program}},
	volume = {70},
    pages = {S8},
	year = {2018}
}

@article{2007APJS.172.1,
	archiveprefix = {arXiv},
	author = {{Capak}, P. and {Aussel}, H. and {Ajiki}, M. and others},
	doi = {10.1086/519081},
	eprint = {0704.2430},
	journal = {\apjs},
	month = {Sep},
	number = {1},
	title = {{The First Release COSMOS Optical and Near-IR Data and Catalog}},
	volume = {172},
    pages = {99},
	year = {2007}
}

@article{2013APJS.208.1,
	archiveprefix = {arXiv},
	author = {{Newman}, Jeffrey and {Cooper}, Michael and {Davis}, Marc and others},
	doi = {10.1088/0067-0049/208/1/5},
	eprint = {1203.3192},
	journal = {\apjs},
	month = {Sep},
	number = {1},
	title = {{The DEEP2 Galaxy Redshift Survey: Design, Observations, Data Reduction, and Redshifts}},
	volume = {208},
    pages = {5},
	year = {2013}
}

@article{2017APJS.233.2,
	archiveprefix = {arXiv},
	author = {{Albareti}, Franco and {Prieto}, Carlos and {Almeida}, Andres and others},
	doi = {10.3847/1538-4365/aa8992},
	eprint = {1608.02013},
	journal = {\apjs},
	month = {Dec},
	number = {2},
	title = {{The 13th Data Release of the Sloan Digital Sky Survey: First Spectroscopic Data from the SDSS-IV Survey Mapping Nearby Galaxies at Apache Point Observatory}},
	volume = {233},
    pages = {25},
	year = {2017}
}

@article{2015AandC.10,
	archiveprefix = {arXiv},
	author = {{Rowe}, Barnaby and {Jarvis}, Michael and {Mandelbaum}, Rachel and others},
	doi = {10.1016/j.ascom.2015.02.002},
	eprint = {1407.7676},
	journal = {A\&C},
	month = {Apr},
	title = {{GalSim: The modular galaxy image simulation toolkit}},
	volume = {10},
    pages = {121},
	year = {2015}
}

@article{1992StatSci.7.4,
	author = {{Gelman}, Andrew and {Rubin}, Donald},
	doi = {10.1214/ss/1177011136},
	journal = {StaSc},
	month = {Nov},
	title = {{Inference from Iterative Simulation Using Multiple Sequences}},
	volume = {7},
	number = {4},
    pages = {457},
	year = {1992}
}

@article{2013MNRAS.429.4,
	author = {{Miller}, L. and {Heymans}, C. and {Kitching}, T. and others},
	doi = {10.1093/mnras/sts454},
	journal = {\mnras},
	month = {Mar},
	number = {4},
	title = {{Bayesian galaxy shape measurement for weak lensing surveys - III. Application to the Canada--France--Hawaii Telescope Lensing Survey}},
	volume = {429},
    pages = {2858},
	year = {2013},
	archiveprefix = {arXiv},
	eprint = {1210.8201}
}

@article{2015ApJ.807.1,
	author = {{Schneider}, M. and {Hogg}, D. and {Marshall}, P. and {Dawson}, W. and {Meyers}, J. and {Bard}, D. and {Lang}, D.},
	doi = {10.1088/0004-637X/807/1/87},
	journal = {\apj},
	month = {Jul},
	number = {1},
	title = {{Hierarchical probabilistic inference of cosmic shear}},
	volume = {807},
    pages = {87},
	year = {2015},
	archiveprefix = {arXiv},
	eprint = {1411.2608}
}

@article{2017MNRAS.466.2,
	author = {{Robotham}, A. and {Taranu}, D. and {Tobar}, R. and {Moffett}, A. and {Driver}, S.},
	doi = {10.1093/mnras/stw3039},
	journal = {\mnras},
	month = {Apr},
	number = {2},
	title = {{PROFIT: Bayesian profile fitting of galaxy images}},
	volume = {466},
    pages = {1513},
	year = {2017},
	archiveprefix = {arXiv},
	eprint = {1611.08586}
}

@article{2023JOSS.8.81,
	author = {{Nightingale}, J. and {Amvrosiadis}, A. and {Hayes}, R. and others},
	doi = {10.21105/joss.04475},
	journal = {JOSS},
	month = {Jan},
	number = {81},
	title = {{PyAutoGalaxy: Open-Source Multiwavelength Galaxy Structure & Morphology}},
	volume = {8},
	year = {2023},
	pages = {4475}
}

@article{2023JOSS.pysersic,
    doi = {10.21105/joss.05703},
    url = {https://doi.org/10.21105/joss.05703},
    year = {2023},
    volume = {8},
    number = {89},
    pages = {5703},
    author = {Pasha, Imad and Miller, Tim B.},
    title = {pysersic: A Python package for determining galaxy structural properties via Bayesian inference, accelerated with jax},
    journal = {JOSS}
}

@article{2023MNRAS.525.1,
	author = {{Rigamonti}, Fabio and {Dotti}, Massimo and {Covino}, Stefano and {Haardt}, Francesco and {Cortese}, Luca and {Landoni}, Marco and {Varisco}, Ludovica},
	doi = {10.1093/mnras/stad2363},
	journal = {\mnras},
	month = {Oct},
	number = {1},
	title = {{Decomposing galaxies with BANG: an automated morphokinematic decomposition of the SDSS-DR17 MaNGA survey}},
	volume = {525},
    pages = {1008},
	year = {2023}
}

@article{2023MNRAS.AstroPhot,
	author = {{Stone}, Connor and {Courteau}, Stéphane and {Cuillandre}, Jean-Charles and {Hezaveh}, Yashar and {Perreault-Levasseur}, Laurence and {Arora}, Nikhil},
	doi = {10.1093/mnras/stad2477},
	journal = {\mnras},
	month = {Aug},
	title = {{AstroPhot: Fitting Everything Everywhere All at Once in Astronomical Images}},
	year = {2023},
    volume = {525},
    pages = {6377},
	archiveprefix = {arXiv},
	eprint = {2308.01957}
}

@article{2006BA.1.1,
	author = {{Blei}, David and {Jordan}, Michael},
	doi = {10.1214/06-BA104},
	journal = {BayAn},
	month = {Mar},
	number = {1},
	title = {{Variational inference for Dirichlet process mixtures}},
	volume = {1},
	year = {2006},
	pages = {121}
}

@article{1995SIAMJSC.16.5,
	author = {{Byrd}, R. and {Lu}, P. and {Nocedal}, J. and {Zhu}, C.},
	doi = {10.1137/0916069},
	journal = {SJSC},
	number = {5},
	title = {{A Limited Memory Algorithm for Bound Constrained Optimization}},
	volume = {16},
	year = {1995},
	pages = {1190}
}

@article{SLSQP,
	author = {{Kraft}, D.},
	journal = {Tech. Rep. DFVLR-FB 88-28, DLR German Aerospace Center – Institute for Flight Mechanics, Koln, Germany},
	title = {{A software package for sequential quadratic programming}},
	year = {1988}
}

@article{2019FTML.12.4,
	author = {{Kingma}, Diederik},
	doi = {10.1561/2200000056},
	journal = {Foundations and Trends in Machine Learning},
	month = {Nov},
	number = {4},
	title = {{An Introduction to Variational Autoencoders}},
	volume = {12},
	year = {2019},
	archiveprefix = {arXiv},
	eprint = {1906.02691},
	pages = {307}
}

@article{2021IEEE.43.11,
	author = {{Kobyzev}, Ivan and {Prince}, Simon and {Brubaker}, Marcus},
	doi = {10.1109/TPAMI.2020.2992934},
	journal = {IEEE Transactions on Pattern Analysis and Machine Intelligence},
	month = {Nov},
	number = {11},
	title = {{Normalizing Flows: An Introduction and Review of Current Methods}},
	volume = {43},
	year = {2021},
	archiveprefix = {arXiv},
	eprint = {1908.09257},
	pages = {3964}
}

@article{2022ApJ.163.4,
	author = {Muyskens, Amanda L. and Goumiri, Imène R. and Priest, Benjamin W. and Schneider, Michael D. and Armstrong, Robert E. and Bernstein, Jason and Dana, Ryan},
	doi = {10.3847/1538-3881/ac4e93},
	journal = {\apj},
	month = {Mar},
	year = {2022},
	volume = {163},
	number = {4},
	title = {{Star–Galaxy Image Separation with Computationally Efficient Gaussian Process Classification}},
	pages = {148},
	archiveprefix = {arXiv},
	eprint = {2105.01106}
}

@article{2022ApJ.924.2,
	author = {{Buchanan}, James and {Schneider}, Michael and {Armstrong}, Robert and {Muyskens}, Amanda and {Priest}, Benjamin and {Dana}, Ryan},
	doi = {10.3847/1538-4357/ac35ca},
	journal = {\apj},
	month = {Jan},
	number = {2},
	title = {{Gaussian Process Classification for Galaxy Blend Identification in LSST}},
	volume = {924},
	year = {2022},
	archiveprefix = {arXiv},
	eprint = {2107.09246},
	pages = {94}
}

@article{2015AandA.575.A41,
    author = {{Guyonnet, A.} and {Astier, P.} and {Antilogus, P.} and {Regnault, N.} and {Doherty, P.}},
    title = {Evidence for self-interaction of charge distribution   in charge-coupled devices},
    DOI= "10.1051/0004-6361/201424897",
    url= "https://doi.org/10.1051/0004-6361/201424897",
    journal = {A\&A},
    year = 2015,
    volume = 575,
    pages = "A41",
    month = "",
    archiveprefix = {arXiv},
    eprint = {1501.01577},
}

@article{2018APJS.234.2,
	archiveprefix = {arXiv},
	author = {{Mao}, Yao-Yuan and {Kovacs}, Eve and {Heitmann}, Katrin and {Uram}, Thomas and {Benson}, Andrew and {Campbell}, Duncan and {Cora}, Sof{\'\i}a and {DeRose}, Joseph and {Di Matteo}, Tiziana and {Habib}, Salman and {Hearin}, Andrew and {Bryce Kalmbach}, J. and {Krughoff}, K. Simon and {Lanusse}, Fran{\c c}ois and {Luki{\'c}}, Zarija and {Mandelbaum}, Rachel and {Newman}, Jeffrey and {Padilla}, Nelson and {Paillas}, Enrique and {Pope}, Adrian and {Ricker}, Paul and {Ruiz}, Andr{\'e}s and {Tenneti}, Ananth and {Vega-Mart{\'\i}nez}, Cristian and {Wechsler}, Risa and {Zhou}, Rongpu and {Zu}, Ying},
	doi = {10.3847/1538-4365/aaa6c3},
	eprint = {arXiv:1709.09665},
	journal = {\apjs},
	month = {Feb},
	number = {2},
	title = {{DESCQA: An Automated Validation Framework for Synthetic Sky Catalogs}},
	volume = {234},
    pages = {36},
	year = {2018}
}
\bibliographystyle{aasjournal}

\appendix

\section{Additional prior distributions}\label{sec:additional_prior}
Section~\ref{sec:prior} details our BGMM prior on four parameters describing the lensed galaxy shape. This appendix describes our prior on angular distributions of galaxy ellipticity and centroid position, and on parameters of the PSF.

\subsection{Angular distributions}\label{sec:angular}
\begin{figure}
\includegraphics[width=\textwidth]{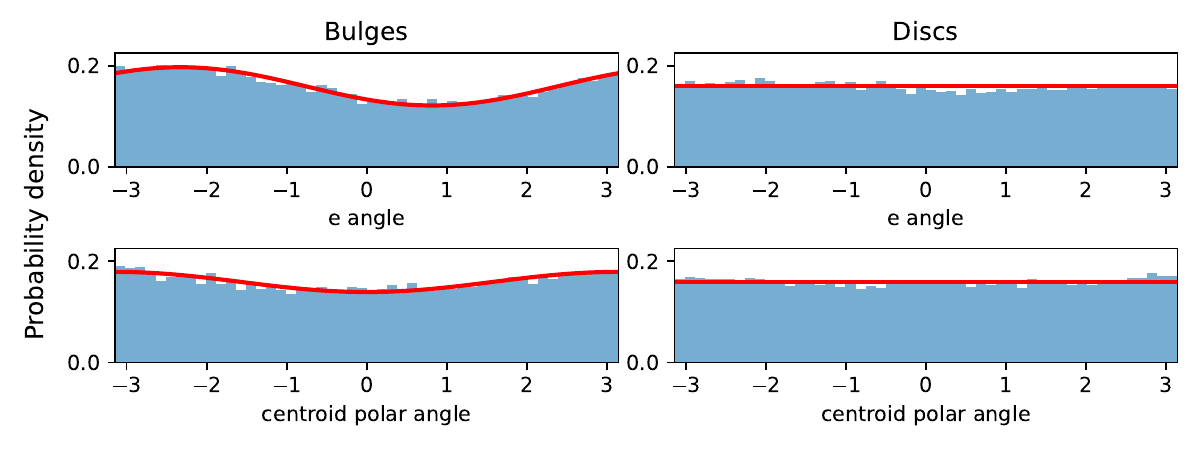}
\caption{Densities for ellipticity angle (top row) and galaxy centroid polar angle around the footprint bounding box center (bottom row), for bulges (left) and disks (right).\label{fig:angular}}
\end{figure}

Figure~\ref{fig:angular} shows the angular distributions for bulges and disks in our dataset (the histograms), overlaid with red curves showing the prior we adopt for each.

As noted in Section~\ref{sec:prior}, while the angular distributions for disks are quite uniform, those for bulges are distinctly sinusoidal. The origin of this effect is unknown, but it may possibly be related to a known bug in the ellipticity distribution used for CosmoDC2~\citep{eve_kovacs_2022_6336066}. Here, we are mainly just concerned with modeling the distribution correctly, regardless of its origin. We model the angular densities for bulges using sinusoids, while we assume constant angular densities for disks. Our prior for bulge ellipticity angle is
\begin{equation}
0.037774 * \sin(\theta + 3.9) + 1 / 2\pi,
\end{equation}
and that for centroid angle is
\begin{equation}
0.0198944 * \cos(\theta + \pi) + 1 / 2\pi,
\end{equation}
for $\theta$ in radians.

\subsection{PSF Distributions}\label{sec:psf}
Section~\ref{sec:parameters} details how we associate ``true'' Kolmogorov PSF parameters (FWHM and ellipticity) to each footprint in our dataset. This appendix describes our prior on these parameters. To simplify the density fit, we omit the most extreme outliers, by restricting our attention only to examples with PSF FWHM, e1, and e2 between the 0.1 and 99.9 percentiles of those values in our dataset.

\begin{figure}
\gridline{\fig{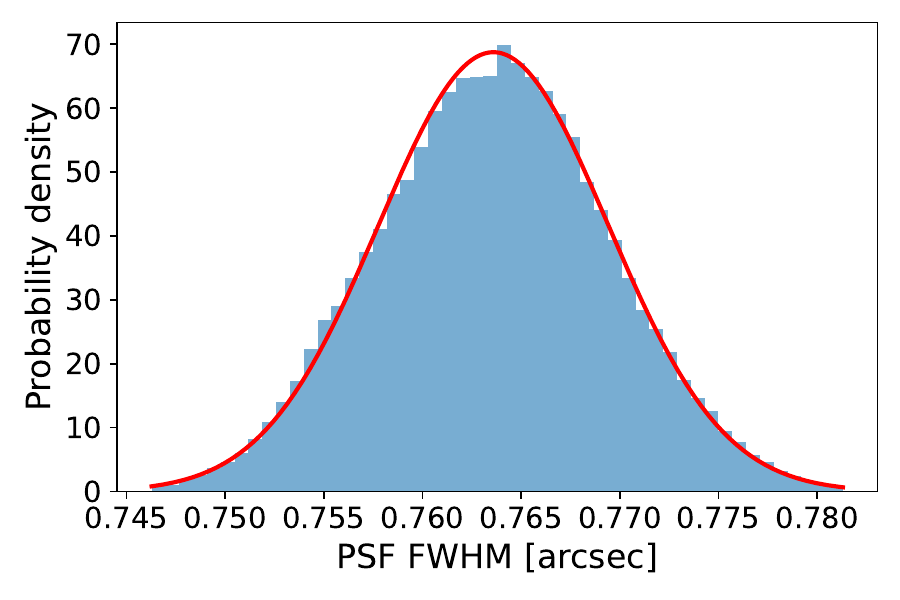}{0.5\textwidth}{(a)}
          \fig{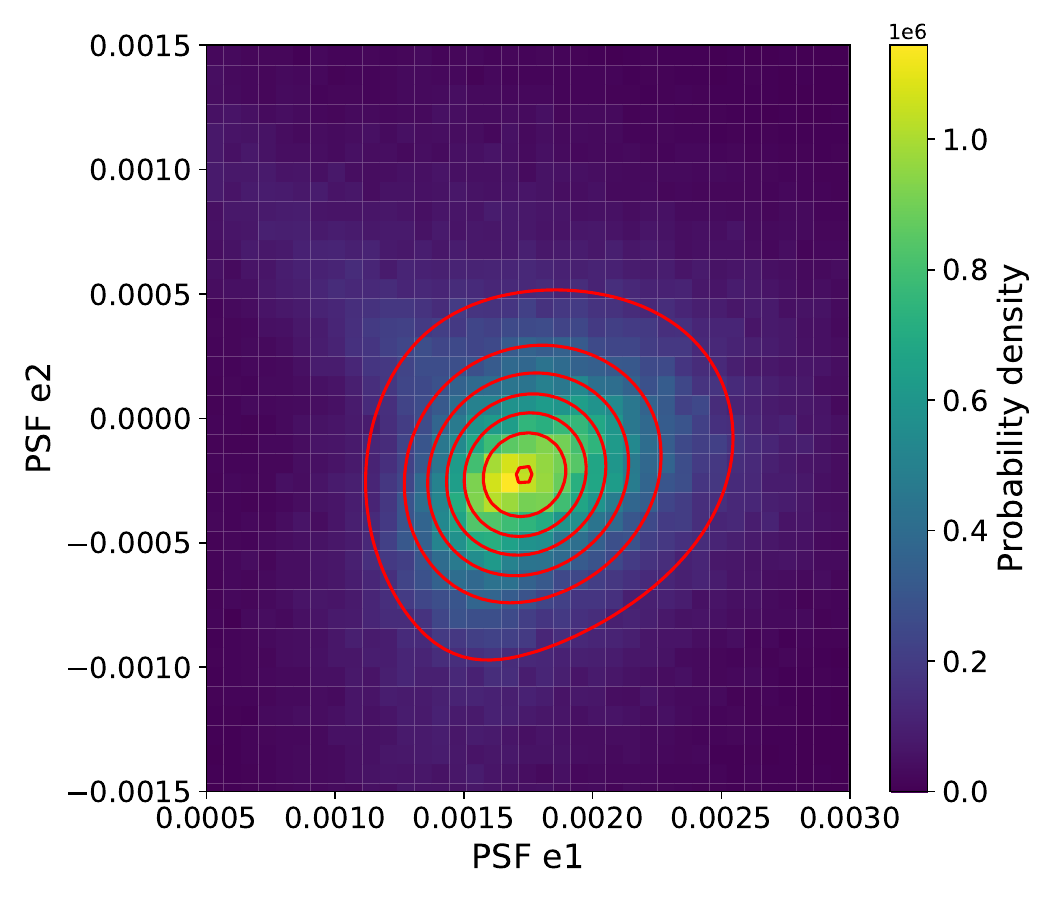}{0.5\textwidth}{(b)}}
\gridline{\fig{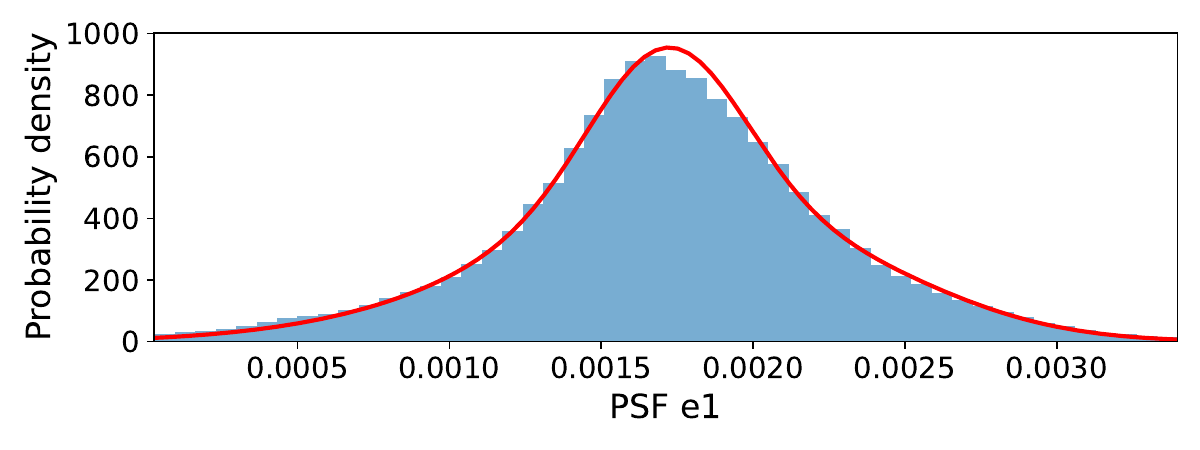}{0.5\textwidth}{(c)}
          \fig{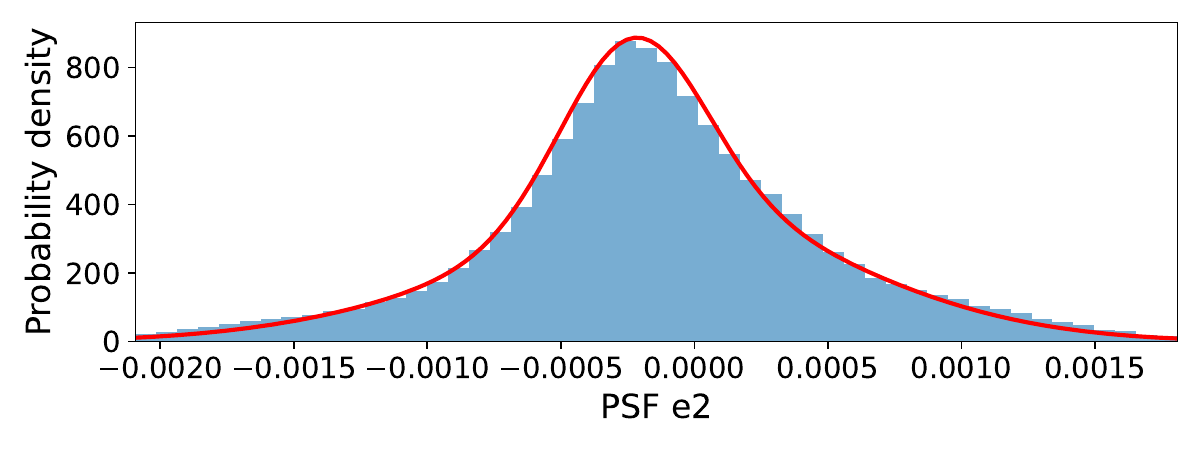}{0.5\textwidth}{(d)}}
\caption{(a) PSF FWHM, (b) PSF e1 and e2 joint distribution, (c) PSF e1 marginal distribution, and (d) PSF e2 marginal distribution. The histograms show the actual distributions of these values in our full dataset, while the red curves show our prior.\label{fig:psf_dists}}
\end{figure}

We do not observe a clear correlation between PSF FWHM and ellipticity, so we fit these distributions separately. As shown in Figure~\ref{fig:psf_dists} (a), PSF FWHM is described very well by a single-component Gaussian distribution with a mean of 0.764\arcsec\ and a standard deviation of 0.00582\arcsec.

For ellipticity, we perform a fit with several 2D Gaussian components. The e1--e2 correlation of each of these components is fixed to 0.0775. We obtain this correlation by an initial fit of a single 2D Gaussian to a highly restricted inner core of our outlier-cleaned data, obtained by selecting only those examples that lie within one standard deviation of the mean for both PSF e1 and e2.

With the correlation thus fixed, the free parameters of each component in the multicomponent fit are its relative weight, its mean in the e1 and e2 axes, and its standard deviation along each of these axes. We performed a maximum-likelihood fit of these values for numbers of components ranging from one through seven. For these fits, we include examples with values of PSF e1 and e2 within three standard deviations of the mean. We observe that four components is the ``elbow'' of the distribution of average likelihood vs.\ number of components. Above this number of components, while the fit likelihood continues to increase, it does so at a markedly slower pace than for the first four. We adopt the four-component fit for our prior, shown in Figures~\ref{fig:psf_dists} (b--d).

\section{Photometric Calibrations and Variance}\label{sec:photometric}
\subsection{Photometric Calibrations}
The DC2 Truth Match catalog gives us the true $i$-band mag for every galaxy, but this does not directly tell us the expected number of instrumental flux units that a galaxy will contribute to a specific exposure or a coadd. Photometric calibrations refer to the relationship between true flux in nJy and the expected number of instrumental flux units. We estimate the photometric calibrations used in this study by comparing the instrumental flux contained in pixels to the true nJy flux that the underlying galaxies contribute to those pixels. We observe that this relationship follows a simple straight line. This appendix describes our calibration process in greater detail.

For each footprint in our dataset, we render an image of the galaxy's true light profile using \texttt{GalSim} and look at the fraction of the total image flux contained in each of the footprint's pixels. For this purpose, the true light profile is taken to be a bulge-plus-disk model with a lensed size, shape, and position given by the DC2 catalogs. The fraction of source flux contained in its footprint (what we call the \textit{footprint aperture correction}) is generally higher than 95\%.

We assume that the instrumental flux produced by the light profile of a typical astrophysical source, appropriately aperture-corrected, has a simple proportional relationship to the source's true nJy flux:
\begin{equation}
I = m * J * a
\end{equation}
where $I$ is the instrumental flux contributed to a footprint from an astrophysical source, $J$ is the source's true overall flux in nJy, $a$ is the footprint aperture correction, and $m$ is an unknown constant of proportionality.

We further assume that the instrumental flux recorded in a pixel is equal to the above astrophysical flux contribution plus a bias, where the bias is constant in each pixel:
\begin{equation}
f = I + b * n
\end{equation}
where $f$ is the total instrumental flux in a footprint, $n$ is the number of pixels in the footprint, and $b$ is a constant per-pixel bias. It follows that
\begin{equation}
f / n = m * (J * a / n) + b.
\end{equation}
For each footprint, we know $f$ and $n$ directly, we know $J$ from the truth catalog, and we compute $a$ as described above. Our photometric calibrations are described by the two unknown parameters $m$ and $b$.

\begin{figure}
\centering
\includegraphics[width=0.5\textwidth]{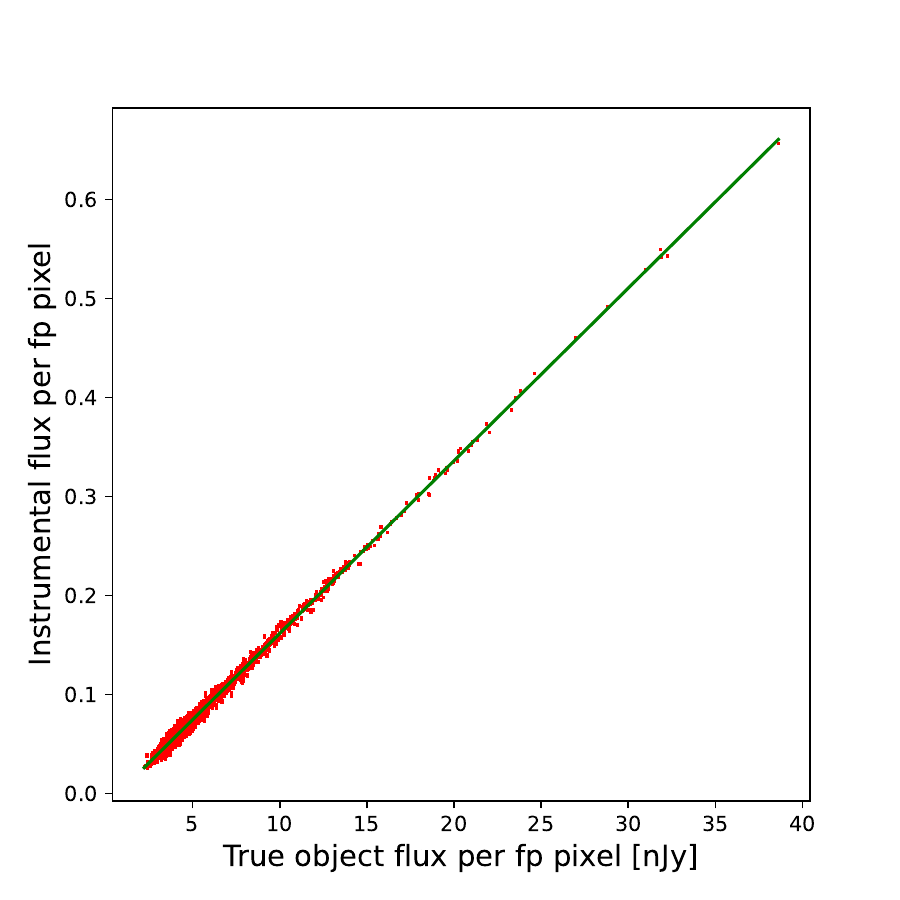}
\caption{Green line: The photometric calibration line we use to define the true instrumental flux for each galaxy. Red dashes: true $i$-band nJy flux per pixel ($x$-axis) vs.\ observed footprint instrumental flux per pixel ($y$-axis) for all galaxies included in this fit. The vertical error bars are derived from the Pipelines-estimated per-pixel variance plane.\label{fig:photometric_calibration}}
\end{figure}

The photometric calibrations describe the mean expected instrumental flux as a function of true nJy flux, while the variance in the instrumental flux in each pixel is directly estimated by the Pipelines. A minimum-chi-square fit of $f / n$ on the $y$-axis vs.\ $J * a / n$ on the $x$-axis should therefore give us the optimal photometric calibrations, described by $m$ and $b$. Figure~\ref{fig:photometric_calibration} shows what this fit looks like on our isolated footprint dataset, after we apply some cuts described below.

As noted in Section~\ref{sec:conditioning_on_selection}, for the realistic dataset, the observed instrumental flux of footprints created from $5\sigma$ galaxies is generally higher than the flux expected to be observed on average from such galaxies, because only observations that pass the $5\sigma$ detection threshold get turned into footprints. If we have the correct mean photometric calibration line, we should not expect to see any footprint fluxes below this line for galaxies with a true brightness at or below the $5\sigma$ level. For sufficiently bright galaxies, the conditional dependence of the likelihood on detection becomes negligible, and the total footprint flux becomes approximately Gaussian-distributed. In practice, we see an approximately Gaussian distribution of residuals around a straight-line fit for galaxies with at least $20\sigma$ brightness in all patches, and we use only these galaxies for our photometric calibrations.

Every patch has photometric calibrations quite similar to every other patch. Some patches contain one or two galaxies vastly brighter than all others and a relatively small number of minimum-$20\sigma$ galaxies overall, diminishing the statistical quality of a straight-line fit to the points in that patch. Hence, for this study, we fit the combined footprints from all patches (with at least $20\sigma$ galaxy brightness) and use the resulting, single photometric calibration line to establish the true expected instrumental flux level for all galaxies across all patches.

To maintain an even distribution of points above and below the calibration line, we place a $20\sigma$ cut on the true galaxy nJy flux, as opposed to instrumental flux. This technically results in a kind of chicken-and-egg problem, because traditionally, when we say $20\sigma$ we are talking about a function of the detection threshold, which is an instrumental flux value; thus, we must convert this to a nJy flux using some kind of photometric calibrations. We therefore start by making an initial crude estimate of the calibration line in each patch by performing a simple least-squares fit of the true galaxy flux on the $y$-axis to the instrumental flux on the $x$-axis (the inverse of our final calibrations). Specifically, we only fit footprints with more than $20\sigma$ instrumental flux and less than $50\sigma$. The upper limit removes a relatively small number of exceptionally bright footprints, which can unduly swing a simple least-squares fit such as this. We declare the $20\sigma$ nJy flux level to correspond to the $20\sigma$ instrumental flux level according to this initial fit.

A handful of outliers show up in all of the straight-line fits described above. The mininum-chi-square fit consults the per-pixel instrumental flux variance estimated from the Pipelines. Based on this pipelines-estimated variance level, we declare any points that sit at least five standard deviations away from the fit line to be outliers. Of the 2175 footprints from across all patches used in the minimum-chi-square fit, 43 are identified as outliers in this way. The fit is then repeated with outliers removed. The final fit, shown in Figure~\ref{fig:photometric_calibration}, has a slope of 0.0174 instrumental flux units per nJy, with a bias of $-0.0126$ instrumental flux units per pixel.

We note that a fit that constraints the bias to be 0 does not reproduce the neat Gaussian distribution of $\chi$ values described in Appendix~\ref{sec:variance}. The negative bias indicates some overall background oversubtraction in the coadded images. If the amount of background oversubtraction depended strongly on the overall flux of an object, we would expect to see pronounced deviations from linearity in the photometric calibration plot. Given the rather high quality of the linear fit (as discussed below, after variance rescaling has been performed), we conclude that the background oversubtraction appears to be relatively constant across all footprints in our dataset. We do not know the origin of this apparent oversubtraction.

\subsection{Instrumental Flux Variance and Covariance}\label{sec:variance}
\begin{figure}
\gridline{\fig{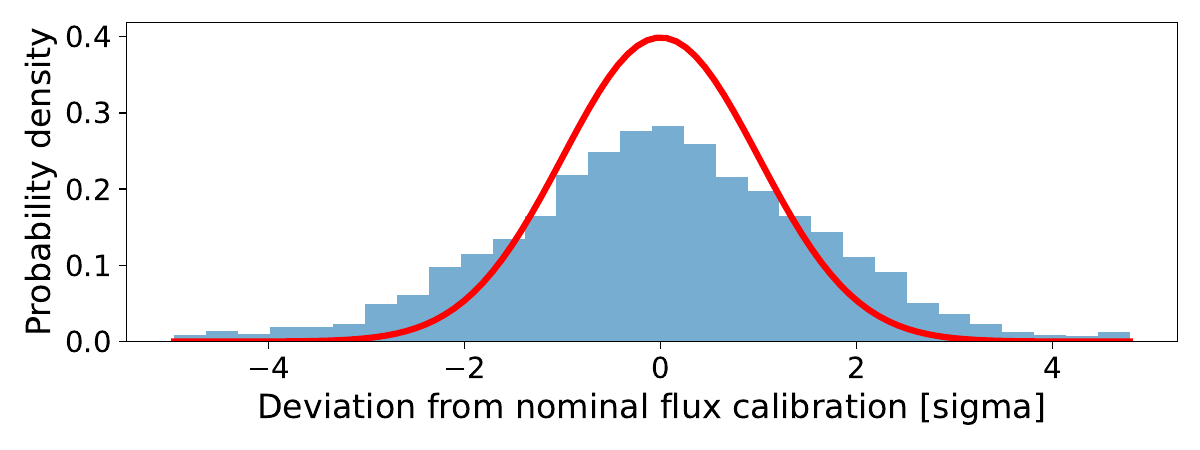}{0.5\textwidth}{(a) Before rescaling}
          \fig{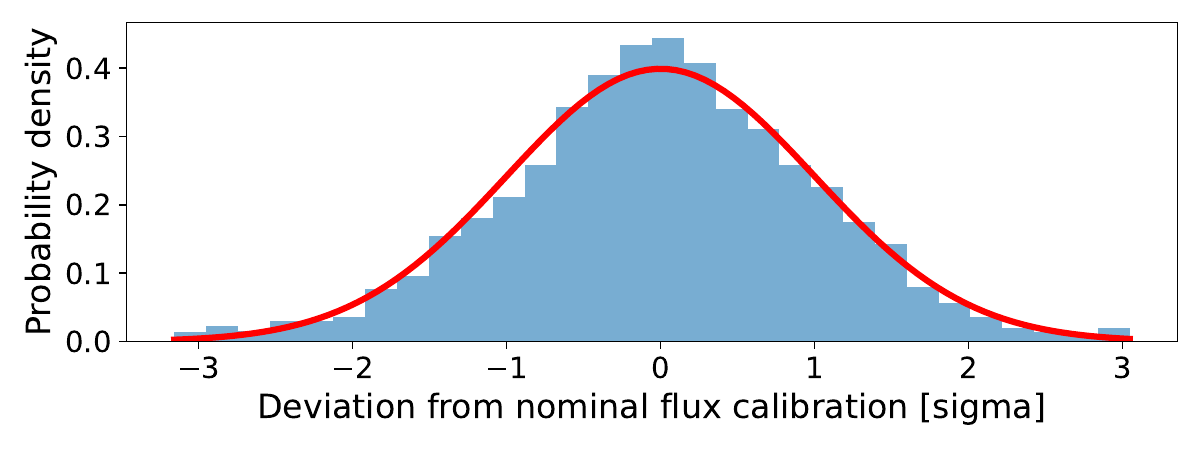}{0.5\textwidth}{(b) After rescaling}}
\caption{Histograms showing the distribution of $\chi$ values for each footprint that contributes to the photometric calibration fit, (a) before and (b) after the Pipelines-supplied variance plane is rescaled, as described in the text. The red curve shows a unit-variance Gaussian distribution centered at zero.\label{fig:chi}}
\end{figure}

If the total instrumental flux is indeed Gaussian-distributed for every galaxy, with a mean given by the true photometric calibration line, and a variance accurately and completely described by the Pipelines-estimated variance plane, then the distribution of the shifted and scaled quantity $\chi$ defined as $(\text{inst flux} - \text{mean}) / \sqrt{\text{variance}}$ should follow a Gaussian distribution with zero mean and unit variance. Figure~\ref{fig:chi} (a) shows what this distribution actually looks like, compared to the theoretical unit-variance distribution in red.

The value of $\sum \chi^{2} / \text{n.d.o.f.\ }$for the calibration fit is 2.469, where the sum of $\chi^{2}$ is taken over all points in the fit, and the effective number of degrees of freedom (n.d.o.f.) is equal to the number of points in the fit minus 2. Correspondingly, the overall distribution of $\chi$-values adheres very well to a unit Gaussian when the variance plane is multiplied by 2.469, as can be seen in Figure~\ref{fig:chi} (b). In other words, the distribution of total instrumental fluxes among the footprints behaves exactly as expected from a simple linear model with Gaussian-distributed errors, but only after the Pipelines-supplied variance plane is scaled up. This may mean that the Pipelines-supplied variance plane is simply too small on average. However, we observe that footprints with a simulated noise level exactly described by the variance plane match the observed level of actual DC2 noise much more closely than ones in which we amplify the noise by a factor of 2.469.

The discrepancy may also indicate that the covariance between pixels is too significant to ignore here. Recall that we model each pixel as totally uncorrelated, and we estimate the variance in a footprint's total instrumental flux by simply adding the separate variances in each pixel. In the more general case of nonzero covariance,
\begin{equation}
\text{Var}\left( \sum_{i} X_{i} \right) = \sum_{i} \text{Var}( X_{i} ) + 2 \sum_{i < j} \text{Cov}(X_{i}, X_{j}).
\end{equation}
Here, $X_{i}$ denotes the instrumental flux in pixel $i$, and the sums range over all the pixels in a given footprint. If the sum of the pairwise covariances between pixels in a typical footprint is equal to 0.735 times the sum of the individual pixel variances, then the variance in the sum of the pixel values will be what we observe above. Nevertheless, in this study, we have not tried to directly estimate the covariance between pixels in a typical footprint.

\section{Approximate Selection Corrections to the Likelihood}\label{sec:detection_correction_appendix}
This appendix extends the discussion in Section~\ref{sec:conditioning_on_selection}. Recall that $D$ is an observed array of instrumental flux values reported in a detection footprint, $\theta$ is some specific vector of our nine model parameters describing the galaxy light profile and PSF, and $det$ is the fact that a footprint for this galaxy ended up in our dataset. To compute the effect of $det$ on the likelihood, we first rewrite the conditional likelihood as
\begin{equation}
\label{eq:conditioned_likelihood}
p(D | \theta, det) = p(D | \theta) p(det | \theta, D) / p(det | \theta)
\end{equation}
using Bayes' theorem. The denominator can be expanded into
\begin{equation}
\label{eq:total_det_prob}
p(det | \theta) = \int dD' p(D' | \theta) p(det | \theta, D')
\end{equation}
in which the integral is taken over all possible image pixel arrays $D'$.

Let $a$ be the condition that the pixel values in a given image pass the Pipelines detection algorithm. Let $b$ be the condition that, among the true objects the footprint pixels overlap, one of those true objects has a flux at least 50 times larger than the total flux of all the others. Let $c$ be the condition that the footprint has exactly one peak. Separating $det$ into the three conditions $a$, $b$, and $c$, equation~\ref{eq:total_det_prob} becomes
\begin{equation}
\begin{aligned}
p(det | \theta, D) &= p(a, b, c | \theta, D) \\
                   &= p(a | \theta, D) p(b | \theta, D, a) p(c | \theta, D, a, b)
\end{aligned}
\end{equation}
For any of the pixel arrays $D$ we consider, $D$ passes the algorithm, and therefore, $p(a | \theta, D)$ is just 1.\footnote{Technically, this is only approximately true, because we define $D$ to only include pixels in the footprint, whereas the PSF-convolved image used for detection includes pixels in the convolution that lie outside the footprint. However, only the central core of pixels in any footprint actually passed the Pipelines detection algorithm (recall that footprints are expanded around the detected pixels by 2.4 times the PSF size), and the center of any footprint is so far away from the edge that pixels outside the footprint make a completely negligible contribution to the PSF convolution at that point.} The conditional detection probability thus simplifies to
\begin{equation}
p(det | \theta, D) = p(b | \theta, D, a) p(c | \theta, D, a, b)
\end{equation}

Now we make some simplifying approximations. First, we assume that detection condition $b$, the fact that a footprint overlaps just one true galaxy of real significance, is for practical purposes independent of $D$ once $\theta$ and $a$ have been specified. We expect that the presence of a secondary, dimmer galaxy in the footprint should not significantly increase the probability of detection unless it lies in a narrow range of brightness values extremely close to the main galaxy (presenting more independent opportunities for detection with comparable likelihood), and/or it overlaps the main galaxy extremely closely along the same line of sight to the telescope (amplifying the detection probability of the regions of overlapping light). We do not rigorously establish the true significance of this interaction here, but it significantly simplifies our analysis to assume total $D$-independence (of $b$, conditioned on $\theta$ and $a$), and it does not seem to introduce obvious pathologies in the results that follow. Under this assumption, the conditional detection probability further simplifies to
\begin{equation}
p(det | \theta, D) = p(b | \theta, a) p(c | \theta, D, a, b)
\end{equation}

Another simplifying approximation is that the probability of condition $c$, that the footprint should contain only one peak, is nearly 100\% for all possible parameter vectors $\theta$ if conditions $a$ and $b$ are both met. We directly observe that this is very nearly true in our dataset---nearly 100\% of data examples for galaxies in the dimmer range, comprising the vast majority our datasets, pass condition $c$ if they first pass conditions $a$ and $b$, and even the brightest galaxies we consider have at least a 90\% chance of doing so. Under the approximation $p(c | \theta, D, a, b) = 1$ for every case, the conditional detection probability finally simplifies to
\begin{equation}
p(det | \theta, D) = p(b | \theta, a)
\end{equation}
for any $\theta$ and for any $D$ likely to be produced from a given $\theta$. Under the same stipulations, equation~\ref{eq:total_det_prob} simplifies to
\begin{equation}
p(det | \theta) = \int_{\mathcal{D}et} dD' p(D' | \theta) p(b | \theta, a)
\end{equation}
where $\mathcal{D}et$ is the set of all pixel value arrays $D'$ that satisfy condition $a$. Since $p(b | \theta, a)$ does not depend on $D'$, it comes out of the integral as a constant. It appears in the denominator of equation~\ref{eq:conditioned_likelihood} for $p(D | \theta, det)$, where it cancels with the equivalent term $p(det | \theta, D)$ in the numerator, leaving
\begin{equation}
p(D | \theta, det) = p(D | \theta) / \int_{\mathcal{D}et} dD' p(D' | \theta)
\end{equation}
as our final expression for the detection-conditioned likelihood.

\end{document}